\documentclass[aps,showpacs,prd,nofootinbib,nobibnotes,notitlepage,twocolumn]{revtex4-2}
\usepackage{graphicx}
\usepackage{amssymb}
\usepackage{amsmath} 
\usepackage{color}
\usepackage{float}
\usepackage{ulem}
\usepackage{accents}
\usepackage{graphicx}
\usepackage{graphicx}
\usepackage{amsfonts}
\usepackage[colorlinks=true,
pdfstartview=FitV,linkcolor=blue,
citecolor=blue,urlcolor=blue,breaklinks=true]{hyperref}
\usepackage{array}
\usepackage{float}
\usepackage{placeins}
\usepackage[dvipsnames]{xcolor}
\usepackage{csquotes}
\usepackage[makeroom]{cancel}
\usepackage{units}
\usepackage{fixmath}
\usepackage{bigints}
\usepackage{amssymb}
\usepackage{systeme}

\newcolumntype{C}[1]{>{\centering\arraybackslash}m{#1}}
\renewcommand{\eqref}[1]{\mbox{Eq.~(\ref{#1})}}

\definecolor{ForestGreen}{rgb}{0.13,0.55,0.13}

\makeatletter
\renewcommand*\l@section{\@dottedtocline{1}{0em}{1.5em}}
\renewcommand*\l@subsection{\@dottedtocline{1}{1.5em}{1.5em}}
\renewcommand*\l@subsubsection{\@dottedtocline{1}{3em}{1.5em}}
\makeatother

\begin{document}

\title{Optical reflection signature of an axion dielectric with magnetic current}

\author{Pedro D. S. Silva$^{a}$}
\email{pedro.dss@ufma.br} \email{pdiegoss.10@gmail.com}
\author{Ronald A. Pereira$^{a}$}\email{ronald123.araujo@gmail.com}
\author{Manoel M. Ferreira Jr.$^{a,b}$}
\email{manojr.ufma@gmail.com}
\affiliation{$^a$Programa de P\'{o}s-gradua\c{c}\~{a}o em F\'{i}sica, Universidade Federal do Maranh\~{a}o, Campus Universit\'{a}rio do Bacanga, S\~{a}o Lu\'is (MA), 65080-805, Brazil}

\affiliation{$^b$Departamento de F\'{\i}sica, Universidade Federal do Maranh\~{a}o,
Campus Universit\'{a}rio do Bacanga, S\~{a}o Lu\'is (MA), 65080-805, Brazil}

\begin{abstract}

In this work, we investigate the reflection properties on the interface between an ordinary dielectric medium and a dielectric supporting a magnetic current (equivalent to a dielectric governed by axion electrodynamics).  Considering the usual Maxwell equations and constitutive relations, we derive the general Fresnel coefficients for reflection for an incident wave with $s$ and $p$ polarization, assuming an isotropic magnetic current on a dielectric substrate. We determine all total internal reflection and critical angles (Brewster angles) conditions, which are given by strict relations between all relevant electromagnetic quantities of the system and the frequency. For $s$- and $p$-polarized incident waves, total internal reflection can occur under certain conditions on the constitutive parameters (for each propagating mode) at specific frequency windows and certain incidence angle intervals. All possible conditions to define critical angles for null reflection are determined.  This scenario allows polarization changes by reflection. Considering a $p$-polarized incident wave, the frequency and the Brewster angle, allowing null reflection for both propagating modes, are determined. The Goos-H\"anchen shift and the complex Kerr rotation are also evaluated. The Kerr ellipticity angle presents a frequency-dependent behavior, also reported in Weyl semimetals, having maximum values $\eta_{Kerr}=\pm \pi/4$ for specific values of frequency, which may work as a signature of this axion chiral dielectric.

\end{abstract}

\pacs{41.20.Jb, 78.20.Ci, 78.20.Fm}
\keywords{Electromagnetic wave propagation; Optical constants;
Magneto-optical effects; Birefringence}

\maketitle


\section{Introduction}

When light impinges on the interface between two distinct media, the reflection and transmission properties are given by the Fresnel coefficients derived from the Maxwell equations, constitutive relations, and boundary conditions\cite{Jackson, Zangwill}. This procedure is also applied for interfaces between non-usual media, such as the ones separating bi-isotropic dielectrics \cite{Sihvola-Fresnel,Hillion} and topological insulators \cite{Ming-Che}, for which the polarization rotation at the Brewster angle may provide a convenient way to determine the magnitude of the chiral coupling \cite{Ming-Che}. A similar investigation in interfaces between topological and left-handed metamaterials revealed the possibility of designing a perfect lens \cite{Zheng-Wei-Zuo}.

Electromagnetic properties of distinct materials can be expressed in terms of optical quantities, such as the Kerr rotation for the reflected wave, the Goos-H\"anchen shift, and birefringence effects. Chiral media, marked by optical activity, are also characterized by the rotatory power (RP), which measures the polarization rotation associated with birefringence \cite{Condon}. The RP is useful to measure the Faraday effect in crystals \cite{Bennett} and plasmas \cite{Porter}, as well as to describe crystals \cite{Dimitriu, Birefringence1},  organic compounds \cite{Barron2, Xing-Liu}, graphene phenomena at terahertz band \cite{Poumirol}, gas of fast-spinning molecules \cite{Tutunnikov}, chiral metamaterials \cite{Woo, Zhang, Mun}, and chiral semimetals \cite{Pesin, Dey-Nandy}. In some particular systems where the RP depends on the frequency, it may undergo sign reversion, which is a feature of rotating plasmas \cite{Gueroult, Gueroult2}, chiral cold plasmas \cite{Filipe1,Filipe2}, and bi-isotropic media with chiral magnetic current \cite{PedroPRB}.

Light reflection on matter surfaces is a useful technique to examine features of distinct materials, including the generation of circularly polarized (CP) and elliptically polarized (EP) waves (not usually available in nature), caused by reflection from materials with a complex refractive index (metals and chiral matter) or total internal reflection. Indeed, total internal reflection from air bubbles in water is one relevant source of EP light in nature \cite{Miller}.

After the incident wave hits the interface of a gyrotropic medium, the reflected wave becomes elliptically polarized, whose polarization characteristic angles (rotation and ellipticity) describe the Kerr rotation \cite{Sato, Argyres, Shinagawa}. In usual scenarios, this rotation happens when the gyration vector depends on the magnetic field, leading to the well-known magneto-optic Kerr effect (MOKE) \cite{Bain}, which is a tool to characterize optical properties of matter systems.

The Kerr rotation has been employed as a useful tool to investigate Weyl semimetals \cite{Kargarian, Ghosh,Sekine}, where the axion term yields novel effects, as frequency-dependent Kerr rotation and ellipticity angles \cite{Trepanier}. These effects can be used to design optical devices, such as chiral filters, circular polarizers or optical isolators, since it is possible to prevent the transmission of RCP and LCP waves at certain frequency ranges \cite{Cote-Trepanier, Cheng-Guo}. In thin-film topological insulators, a giant Kerr rotation of $\pi/2$ in the quantized limit has been predicted to occur \cite{Wang-Kong-Tse} at a low-frequency regime.  The magnitude of Kerr rotations is quite small, being less than 1 degree in usual materials \cite{Schlenker-Souche} and of the order $10^{-6}$ to $10^{-4}$ rad in topological insulators \cite{Sonowal}. The evaluation of the Kerr rotation has been magenta used as a relevant tool to probe different materials,  high-precision measurement in quantum systems \cite{Tong-Li}, unconventional superconductors \cite{Kapitulnik}, and media modeled by CPT-even and Lorentz symmetry violating electrodynamics \cite{Ruiz-Escobar}.

Another scenario studied in the last few years is dielectric matter supporting magnetic currents, in connection with the chiral magnetic effect (CME). Such an effect consists of a macroscopic magnetic current law, ${\bf{J}}= \sigma^{B} {\bf{B}}$, originated from the asymmetry between the number density of left- and right-handed chiral fermions \cite{Kharzeev1, Fukushima}. The CME has been investigated in several distinct situations \cite{Schober, Vilenkin,Maxim, Maxim1, Leite, Dvornikov, Maxim1}, including Weyl semimetals (WSMs) \cite{Burkov, Barnes, Xiaochun-Huang}. Propagation of electromagnetic waves in a dispersive dielectric endowed with magnetic conductivity, assumed as an intrinsic property of the medium on equal footing to the permittivity, was also scrutinized revealing unusual features \cite{Pedro1,PedroPRB2024A,PedroPRB2024B}.

 The vast scenario of research on the relevance of electromagnetic properties for optical material characterization, as well as recent developments in the electrodynamics of dielectrics bearing magnetic currents (in connection with WSMs), is a strong motivation for investigating the optical reflection properties in an axion coupling dielectric surface. In this sense, the present work is devoted to analyzing the optical reflection aspects in an interface between a usual dielectric and a dielectric endowed with magnetic current, ${\bf{J}}= \sigma^{B} {\bf{B}}$, where $\sigma^{B}$ is the magnetic conductivity. 

This paper is outlined as follows. In Sec.~\ref{BasicsMedium}, we briefly review the basic aspects of dielectric media endowed with magnetic current. In Sec.~\ref{reflection-coefficients-section}, we derive the reflection coefficients considering incident wave with $s$- and $p$-polarizations, examining the exotic frequency dependence on the total reflection. In Sec.~\ref{Critical-angles-section}, we address the critical angles for null reflection that lead to the determination of the Brewster angles. In Sec.~\ref{other-optical-effects-section}, other optical properties, such as Goos-H\"anchen effect and complex Kerr rotation, are discussed. Finally, we summarize our results and prospects in Sec. \ref{Final-Remarks-section}.

\section{\label{BasicsMedium}Chiral dielectric endowed with magnetic conductivity }

Classical electromagnetic and optical properties of a dielectric medium in the presence of a magnetic current density have been recently addressed \cite{Pedro1,PedroPRB2024A,PedroPRB2024B}. In this section, we briefly review the main aspects, taking as a starting point the Maxwell equations,
\begin{subequations}
	\begin{eqnarray}
		\nabla\cdot{\bf D} \!&=&\! 0 \; ,
		\hspace{0.3cm}  \hspace{0.3cm}
		\nabla\times{\bf E}+\frac{\partial{\bf B}}{\partial t} = {\bf 0} \; ,
		\label{eqdivD}
		\\
		\nabla\cdot{\bf B} \!&=&\!0 \; ,
		\hspace{0.3cm}  \hspace{0.3cm}
		\nabla\times{\bf H}-\frac{\partial{\bf D}}{\partial t} = {\bf J} \; ,
		\label{eqdivB}
	\end{eqnarray}
\end{subequations}
where ${\bf J}$ is the current density. The system is ruled by conventional constitutive relations, 
\begin{equation}
{\bf{D}}= \epsilon {\bf{E}},  \, \, \, {\bf{B}}= \mu {\bf{H}},
\label{CR1}
\end{equation}
where $\epsilon$ is the electric permittivity and $\mu$ is the magnetic permeability. 
The magnetic current constitutive relation can be written in terms of the conductivity tensor, $\sigma^{B}_{ij}$, 
\begin{equation}
	{J}^{i}={\sigma}_{ij}^{B} \, {B}^{j} \; 
	\label{JCMEij}
\end{equation}
which replaced in Maxwell´s equations, in the plane-wave \textit{ansatz}, yields
\begin{equation}
	({\bf{k}}\times {\bf{k}}\times {\bf{E}})^{i} +  \mu \, \epsilon \, \omega^2 \, {{E}^{i}}
	+ \mathrm{i} \, \mu \, {\sigma}_{ij}^{B} ({\bf{k}} \times {\bf{E}})^{j}  =0\,.
	\label{Ampere1b}
\end{equation}

The wave equation for the electric field amplitude is
\begin{align}\label{EqwaveE0}
	\left[ \, {\bf k}^2 \, \delta_{ij} - k_{i}\,k_{j}-\omega^2 \mu \, \overline{\epsilon}_{ij}(\omega) \, \right] E^{j}=0 \; ,
\end{align}
with the effective permittivity tensor
\begin{eqnarray}
	\overline{\epsilon}_{ij}(\omega)= \epsilon(\omega) \, \delta_{ij}
	-\frac{i}{\omega^2} \, \sigma^{B}_{ia} \, \epsilon_{abj} \, k_{b} \; .
	\label{permittivityCME}
\end{eqnarray}
For an isotropic magnetic conductivity, $\sigma^{B}_{ij}=\Sigma\,\delta_{ij}$, where $\Sigma$ is a real parameter, one obtains a Hermitian permittivity,
\begin{equation}
	{\bar{\epsilon}}_{ij}(\omega )=\epsilon \, {\delta }_{ij}+{\frac{{\mathrm{i}\Sigma }}{{\omega }^{2}}} \, \epsilon_{ijb} \, k_{b} \, ,
	\label{eq61a}
\end{equation}
which assures energy conservation and the absence of dissipation. Relation (\ref{EqwaveE0}) may also be written as
\begin{equation}
	{M}_{ij}E^{j}=0,
	\label{Mequation}
\end{equation}
with
\begin{equation}
	{M}_{ij}= {\bf k}^2 \, \delta_{ij} - k_{i}\,k_{j}-\omega^2 \mu \, \overline{\epsilon}_{ij}(\omega) \; .
	\label{MijCME}
\end{equation}
The dispersion relations of the model are the non-trivial solutions of \eqref{Mequation}, obtained requiring $\det[{ M_{ij} }] = 0$, which provides distinct real refractive indices,
\begin{equation}
	\label{n23-0-1}
		n_{\pm}=\sqrt{\mu\epsilon+\left(\frac{\mu\Sigma}{2\omega}\right)^2}\pm \frac{\mu\Sigma}{2\omega} \; .
\end{equation}
These indices were first carried out in Refs. \cite{Pedro1,PedroPRB2024A}, defining a dispersive non-absorbing medium endowed with birefringence. The propagation modes for the indices (\ref{n23-0-1}), obtained from the relation $M_{ij}E_{j}=0$, are
\begin{equation}
	\label{eigenvector2}
	\mathbf{E}_{\pm}=\frac{1}{n\sqrt{2(n_1^2+n_3^2)}}
	\begin{pmatrix}
		n n_3 \mp \mathrm{i}n_1 n_2 \\
		\pm\mathrm{i}(n_1^2+n_3^2) \\
		\mp\mathrm{i}n_2 n_3-n n_1 \\
	\end{pmatrix} \; ,
\end{equation}
which for a wave propagating at the ${\cal Z}$-axis,
$\mathbf{n}=(0,0,n_{3})$,  yields
\begin{equation}
	\label{eq:polarizations-isotropic}
	\mathbf{E}_{\pm}=\frac{1}{\sqrt{2}}\begin{pmatrix}
		1 \\
		\pm\mathrm{i} \\
		0 \\
	\end{pmatrix}\,.
\end{equation}
The circular birefringence  is quantified by the specific rotatory power $\delta$ \cite{Fowles, Pedro-2021, Hecht}, 
\begin{equation}
	\label{eq:rotatory-power1}
	\delta=-\frac{\omega}{2}[\mathrm{Re}(n_{+})-\mathrm{Re}(n_{-})] \; ,
\end{equation}
which measures the rotation of the oscillation plane of linearly polarized light per unit traversed length in the medium, with $n_{+}$ and $n_{-}$ associated with left and right-handed circularly polarized waves, respectively. For the refractive indices (\ref{n23-0-1}), the rotatory power is 
\begin{equation}\label{eq:rotatory-power}
	\delta=-\frac{\mu\Sigma}{2} \; .
\end{equation}

There exists a relevant interplay between the CME and the Maxwell-Carroll-Field-Jackiw (MCFJ) electrodynamics \cite{CFJ,Colladay}, which contains non-dynamical axion terms. In fact, the magnetic current term appears in the axion Lagrangian \cite{KDeng,Wilczek, Sekine, Tobar,Qiu},
\begin{equation}
	\mathcal{L}=-\frac{1}{4}F^{\mu\nu}F_{\mu\nu}+\theta (\mathbf{E}\cdot \mathbf{B)},
\end{equation}%
where the axion field $\theta$ modifies the non-homogeneous Maxwell equations with derivative terms, 
\begin{align}
	\mathbf{\nabla }\cdot \mathbf{E}&=\rho -\mathbf{\nabla }\theta \cdot \mathbf{B%
	}, \label{M1A} \\
	\mathbf{\nabla }\times \mathbf{B}-\partial _{t}\mathbf{E}&=%
	\mathbf{j}+(\partial _{t}\theta )\mathbf{B}+\mathbf{\nabla }\theta \times 
	\mathbf{E}.  \label{M1B}
\end{align} 
The axion gradient composes the anomalous charge density term, $\mathbf{\nabla }\theta \cdot \mathbf{B}$, and the anomalous Hall current, $\mathbf{\nabla }\theta \times \mathbf{B}$. On the other hand, the time derivative, $(\partial _{t}\theta )\mathbf{B}$, plays the role of the chiral magnetic current \cite{Qiu}. For a cold axion dark matter, $\mathbf{\nabla }\theta
={\bf{0}}$, equations (\ref{M1A}) and (\ref{M1B}) read
\begin{equation}
	\mathbf{\nabla }\cdot \mathbf{E}=\rho, \quad
	\mathbf{\nabla }\times 
	\mathbf{B}-\partial _{t}\mathbf{E}=\mathbf{j}+(\partial _{t}\theta )\mathbf{B},
	\label{Maxwellaxion1}
\end{equation}
with $(\partial _{t}\theta )$ standing in the place of the chiral magnetic conductivity. The connection between CME and axion electrodynamics finds applications in Weyl semimetals \cite{Gomez,Ruiz2,Marco} and Casimir effect \cite{Ruiz1}. Nonlocal extensions of axion electrodynamics were implemented in extended constitutive relations to investigate optical properties of exotic metamaterials \cite{Barredo}.

\section{\label{reflection-coefficients-section}Reflection at the interface between ordinary dielectric and dielectric endowed with magnetic conductivity }

\begin{figure}[h]
\centering\includegraphics[scale=.85]{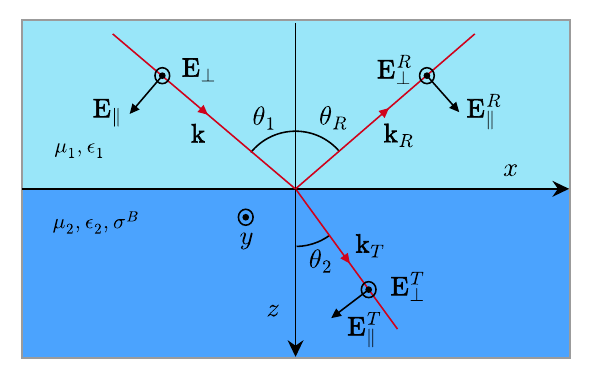}\caption{Interface between a usual and a chiral dielectric. Electric field components of the incident, reflected, and transmitted waves.}%
\label{figura-figura-campos-incidente-refletido-transmitido}
\end{figure}

Let us consider a scenario of reflection of the electromagnetic wave at the interface between two media: a conventional dielectric $(\mu_{1}, \epsilon_{1})$, and a dielectric endowed with magnetic conductivity $(\mu_{2}, \epsilon_{2}, \sigma^{B})$. More specifically, we consider the case where the medium 2 has isotropic magnetic conductivity, $\sigma^{B}_{ij} = \Sigma \, \delta_{ij}$, for which the refractive index of the medium 2 is given by \eqref{n23-0-1}, that is, 
	\begin{align}
		n_{2 \pm} &= \sqrt{\mu_{2} \epsilon_{2} + \left( \frac{ \mu_{2} \Sigma}{2\omega} \right)^{2}} \pm \frac{\mu_{2} \Sigma}{2\omega} , \label{reflection-with-magnetic-conductivity-5}
\end{align}
as discussed in Sec.~\ref{BasicsMedium}. Assuming that both media are governed by conventional constitutive relations as the ones of \eqref{CR1} and taking the usual boundary conditions for the electromagnetic field in the medium with magnetic conductivity, the reflection equations for the fields of Fig. \ref{figura-figura-campos-incidente-refletido-transmitido} read
\begin{align}
E^{R}_{\perp} &= r_{\perp}  E_{\perp}, \label{reflection-with-magnetic-conductivity-1} \\
E^{R}_{\parallel} &= r_{\parallel}  E_{\parallel}, \label{reflection-with-magnetic-conductivity-2} 
\end{align}
where the Fresnel coefficients are 
\begin{align}
r_{s} &= r_{\perp} = \frac{  n_{1} \mu_{2} \cos\theta_{1} - n_{2} \mu_{1} \cos \theta_{2}} { n_{1} \mu_{2} \cos\theta_{1}  + n_{2} \mu_{1} \cos \theta_{2} } , \label{reflection-with-magnetic-conductivity-3} \\
r_{p} &= r_{\parallel} = \frac{  n_{2} \mu_{1} \cos\theta_{1} - n_{1} \mu_{2} \cos\theta_{2}}{n_{2} \mu_{1} \cos\theta_{1} + n_{1} \mu_{2} \cos\theta_{2} } . \label{reflection-with-magnetic-conductivity-4}
\end{align}

Taking into account that medium 2 presents two different refractive indices, see $n_{2+}$ and $n_{2-}$ of \eqref{reflection-with-magnetic-conductivity-5}, each mode travels with different phase velocities and there occurs two refracted (or transmitted) waves \cite{Sihvola-Fresnel}, each one associated with one propagating mode. This is a manifestation of birefringence, a feature of optical active media.

Considering $n_{1} =\sqrt{\mu_{1} \epsilon_{1}}$ and the indices (\ref{reflection-with-magnetic-conductivity-5}), the Fresnel coefficients for polarization-s and polarization-p are
\begin{align}
r_{\perp}^{\pm} &= \frac{ f_{\theta_{1}}- \mu_{1} \sqrt{ \mu_{2} \epsilon_{2} + \left( \frac{ \mu_{2} \Sigma}{2\omega} \right)^{2}} \cos\theta_{2} \mp \frac{ \mu_{2} \Sigma}{2\omega} \mu_{1} \cos\theta_{2} } {    f_{\theta_{1}} + \mu_{1} \sqrt{ \mu_{2} \epsilon_{2} + \left( \frac{ \mu_{2} \Sigma}{2\omega} \right)^{2}} \cos\theta_{2} \pm \frac{ \mu_{2} \Sigma}{2\omega} \mu_{1} \cos\theta_{2} } , \label{reflection-with-magnetic-conductivity-6}
\end{align}
\begin{align}
r_{\parallel}^{\pm} &= \frac{\mu_{1} \sqrt{ \mu_{2} \epsilon_{2} + \left( \frac{ \mu_{2} \Sigma}{2\omega} \right)^{2}} \cos\theta_{1} \pm \frac{ \mu_{2}\Sigma}{2\omega} \mu_{1} \cos\theta_{1}  -f_{\theta_{2}} }  {    \mu_{1} \sqrt{ \mu_{2} \epsilon_{2} + \left( \frac{ \mu_{2} \Sigma}{2\omega} \right)^{2}} \cos\theta_{1} \pm \frac{ \mu_{2}\Sigma}{2\omega} \mu_{1} \cos\theta_{1}  + f_{\theta_{2}}                       }  ,  \label{reflection-with-magnetic-conductivity-7}
\end{align}
with
\begin{align}
f_{\theta_{1,2}}= \mu_{2} \sqrt{\mu_{1} \epsilon_{1} } \cos \theta_{1,2} . \label{reflection-with-magnetic-conductivity-7-1}
\end{align}
Given the coefficients (\ref{reflection-with-magnetic-conductivity-6})  and (\ref{reflection-with-magnetic-conductivity-7}), one can write the reflection coefficient $R$ (or reflectance) as follows \cite{Zangwill}:
\begin{align}
R^{\pm}_{\perp} &= \left|  r_{\perp}^{\pm} \right|^{2} , \label{reflection-with-magnetic-conductivity-8} \\
R^{\pm}_{\parallel} &= |  r_{\parallel}^{\pm}  |^{2} . \label{reflection-with-magnetic-conductivity-9}
\end{align}

Unlike the usual dielectric scenario (where $\Sigma=0$), the $R$ coefficients (\ref{reflection-with-magnetic-conductivity-8}) and (\ref{reflection-with-magnetic-conductivity-9}) depend on the frequency of the incident electromagnetic wave, $\omega$, and on the incidence and refraction angles. This kind of dependence appears typically in interfaces separating an ordinary dielectric and a conducting medium. The coefficients (\ref{reflection-with-magnetic-conductivity-8}) and (\ref{reflection-with-magnetic-conductivity-9}) can be rewritten in terms of the transmission angle $\theta_{2}$ by using the Snell law, $n_{1} \sin \theta_{1} = n_{2 \pm} \sin \theta_{2\pm}$, implying
\begin{align}
\theta_{2\pm} &= \mathrm{arc} \sin \left[   \frac{ \sqrt{\mu_{1}\epsilon_{1}}}{ \sqrt{\mu_{2} \epsilon_{2} + \left( \frac{ \mu_{2} \Sigma}{2\omega} \right)^{2}} \pm \frac{ \mu_{2} \Sigma}{2\omega} }  \sin \theta_{1} \right] . \label{reflection-with-magnetic-conductivity-12}
\end{align}
The latter expressions are obtained assuming that both modes obey Snell's law \cite{Sihvola-Fresnel, Sihvola-Lindell-Fresnel}. This scenario also happens for both ordinary and extraordinary rays in anisotropic materials when incident light impinges the interface perpendicularly to the optical axis \cite{Jun-Fang-Wu, Kumar-Ghatak}.

To get more physical insights about the influence of $\Sigma$ on $R^{\pm}_{\perp} $ and $R^{\pm}_{\parallel}$, we will plot them in terms of frequency for $\theta_{1} (rad)= \displaystyle\{0, \frac{\pi}{6}, \frac{\pi}{4},  \frac{\pi}{3}, \frac{5\pi}{12} \}$.

\subsection{Incident wave with $s$-polarization}

\begin{figure}[h]
	\centering\includegraphics[scale=.69]{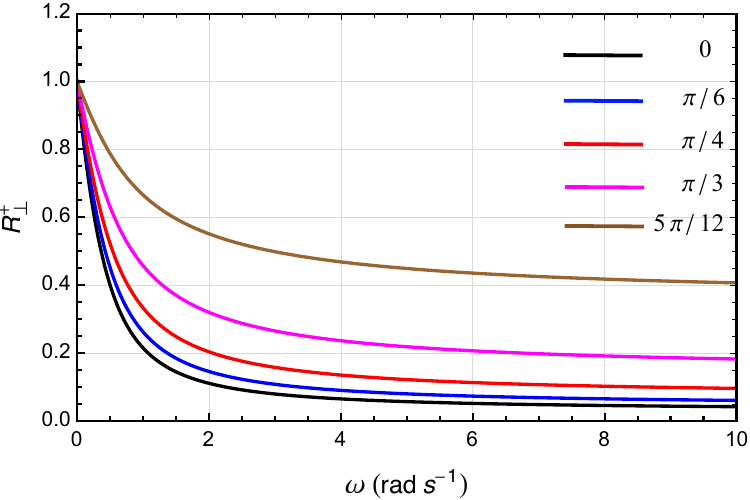}\caption{Reflection coefficient $R^{+}_{\perp}$ in terms of frequency $\omega$ for fixed incidence angles. For the medium 1, we have used $\mu_{1}=1$, $\epsilon_{1}=1$, while for the medium 2: $\mu_{2}=1$, $\epsilon_{2}=2$, $\Sigma = 2$ $\mathrm{s}^{-1}$.}%
	\label{plot-R-perpendicular-mais-angulos-fixos}
\end{figure}

For the incident wave in $s$-polarization, the reflection coefficient is given by the two possibilities, $R^{\pm}_{\perp}$, given in \eqref{reflection-with-magnetic-conductivity-8}. We begin analyzing the behavior of the coefficient $R^{+}_{\perp}$, whose magnitude decreases with the frequency, as depicted in Fig. \ref{plot-R-perpendicular-mais-angulos-fixos}. 

\begin{figure}[h]
\centering\includegraphics[scale=.69]{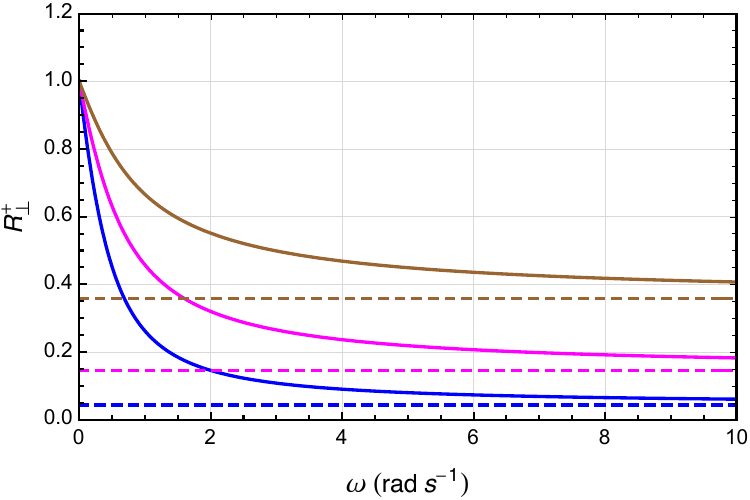}\caption{Reflection coefficient $R^{+}_{\perp}$ in terms of frequency $\omega$ for fixed incidence angles $\theta_{1}=\pi/6$ (blue), $\theta_{1}=\pi/3$ (magenta), and $\theta_{1}=5\pi/12$ (brown). For the medium 1, we have used $\mu_{1}=1$, $\epsilon_{1}=1$, and for the medium 2: $\mu_{2}=1$, $\epsilon_{2}=2$, $\Sigma = 2$ $\mathrm{s}^{-1}$ (solid curves), and $\Sigma=0$ (dashed lines). }%
\label{plot-R-perpendicular-mais-angulos-fixos-with-USUAL-cases}%
\end{figure}

As shown, the asymptotic value of $R^{+}_{\perp}$ (defined for very large frequencies) diminishes with the incidence angle. In true, this happens equally for the  asymptotic value of $R^{-}_{\perp}$, since the expressions (\ref{reflection-with-magnetic-conductivity-6}) and (\ref{reflection-with-magnetic-conductivity-8}) yield the same result in the high-frequency limit, namely
\begin{align}
R^{\pm}_{\perp} \simeq  \left|  \frac{ \mu_{2} \sqrt{\mu_{1} \epsilon_{1}} \cos \theta_{1} - \mu_{1} \sqrt{\mu_{2} \epsilon_{2}} \cos\theta_{2+} }  { \mu_{2} \sqrt{\mu_{1} \epsilon_{1}} \cos \theta_{1} + \mu_{1} \sqrt{\mu_{2} \epsilon_{2} } \cos \theta_{2+} }  \right|^{2} , \label{reflection-with-magnetic-conductivity-17}
\end{align}
or
\begin{align}
R^{\pm}_{\perp} \simeq  \left|    \frac{    \displaystyle \mu_{2}  \sqrt{\mu_{1} \epsilon_{1} } \cos \theta_{1} - \mu_{1} \sqrt{\mu_{2} \epsilon_{2}}  \sqrt{1 - \frac{ \mu_{1} \epsilon_{1} }{ \mu_{2} \epsilon_{2}} \sin^{2} \theta_{1} }  }      {   \displaystyle       \mu_{2} \sqrt{\mu_{1} \epsilon_{1} }  \cos\theta_{1} + \mu_{1}  \sqrt{ \mu_{2} \epsilon_{2}}   \sqrt{ 1 - \frac{ \mu_{1}\epsilon_{1}} {\mu_{2} \epsilon_{2}} \sin^{2} \theta_{1}}}  \right|^{2}. \label{reflection-with-magnetic-conductivity-18}
\end{align}
The asymptotic expression (\ref{reflection-with-magnetic-conductivity-18}) reveals that the effect of the magnetic conductivity is suppressed in the high-frequency limit, indicating that $R_{\perp}^{\pm}$ recover the reflection coefficient usual case (obtained for dielectric with $\Sigma=0$). This feature is properly illustrated in Fig.~\ref{plot-R-perpendicular-mais-angulos-fixos-with-USUAL-cases} for some parameter values.

\begin{figure}[h]
\centering\includegraphics[scale=.69]{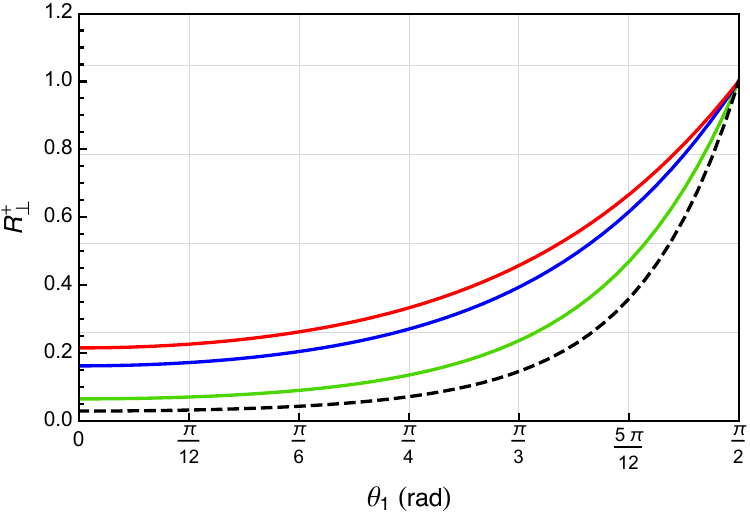}\caption{Reflection coefficient $R^{+}_{\perp}$ in terms of the incidence angle $\theta_{1}$. For the medium 1, we have used $\mu_{1}=1$, $\epsilon_{1}=1$, and for the medium 2: $\mu_{2}=1$, $\epsilon_{2}=2$, $\Sigma/\omega = 0.5$ (green), $\Sigma/\omega = 1.5$ (blue), and $\Sigma/\omega = 2$ (red). The dashed black line represents the usual case ($\Sigma=0$).} 
\label{plot-R-perpendicular-mais-frequencia-fixa}
\end{figure}

The dependence of $R^{+}_{\perp}$ on the incidence angle $\theta_{1}$ is depicted in Fig.~\ref{plot-R-perpendicular-mais-frequencia-fixa}. We observe that by increasing the value of $\Sigma$ (in terms of $\Sigma/\omega$), the $R^{+}_{\perp}$ curve moves away from the standard scenario (dashed black line), enhancing the magnitude of the reflection coefficient.

\begin{figure}[h]
\centering\includegraphics[scale=.69]{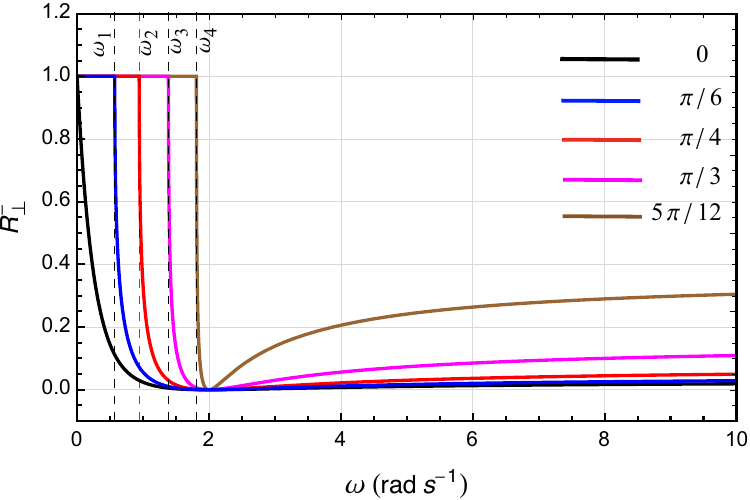}\caption{Reflection coefficient $R^{-}_{\perp}$ in terms of frequency $\omega$ for fixed incidence angles. For the medium 1 we have used $\mu_{1}=1$, $\epsilon_{1}=1$, and for the medium 2: $\mu_{2}=1$, $\epsilon_{2}=2$, $\Sigma = 2$ $\mathrm{s}^{-1}$. The vertical dashed lines indicate the frequencies $\omega_{i}$ of \eqref{reflection-with-magnetic-conductivity-14} which define the limit of the total reflection window (for each case).}%
\label{plot-R-perpendicular-menos-angulos-fixos}%
\end{figure}

As for the reflection coefficient $R^{-}_{\perp}$, its behavior in terms of frequency differs appreciably from the one of $R^{+}_{\perp}$, as shown in Fig. \ref{plot-R-perpendicular-menos-angulos-fixos}. For a non-null $\Sigma$ and general incidence angles (excluding normal incidence), the reflection coefficient becomes equal to 1 in the low-frequency region defined by $0< \omega \leq \omega_{i}$, where $\omega_{i}$ is given by
\begin{align}
	\omega_{i} &= \frac{ \mu_{2} \,\Sigma \, \sqrt{\mu_{1}\epsilon_{1} } \, \left| \sin \theta_{1}  \right|  } { \left|  \mu_{2} \epsilon_{2} - \mu_{1} \epsilon_{1} \sin^{2} \theta_{1} \right| } . \label{reflection-with-magnetic-conductivity-14}
\end{align}

\begin{figure}[h]
\centering\includegraphics[scale=.69]{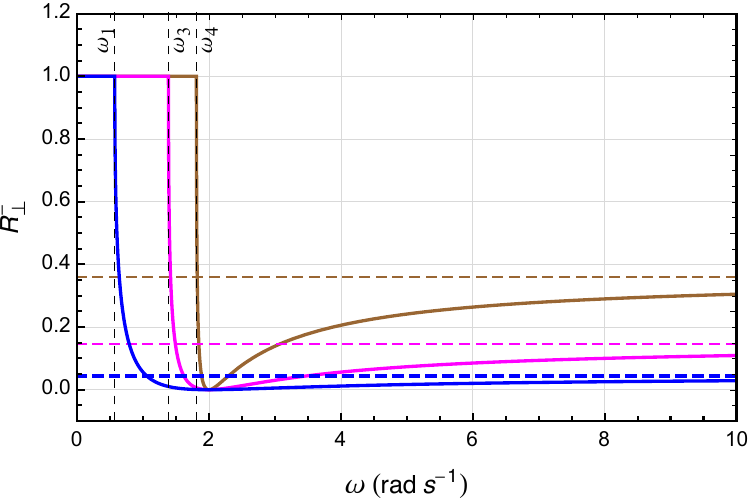}\caption{Reflection coefficient $R^{-}_{\perp}$ in terms of frequency $\omega$ for fixed incidence angles $\theta_{1}=\pi/6$ (blue), $\theta_{1}=\pi/3$ (magenta), and $\theta_{1}=5\pi/12$ (brown). For the medium 1, we have used $\mu_{1}=1$, $\epsilon_{1}=1$, while for the medium 2: $\mu_{2}=1$, $\epsilon_{2}=2$, $\Sigma = 2$ $\mathrm{s}^{-1}$ (solid curves), and $\Sigma=0$ (dashed lines). }%
\label{plot-R-perpendicular-menos-angulos-fixos-with-USUAL-cases}%
\end{figure}
\begin{figure}[h]
	\centering\includegraphics[scale=.69]{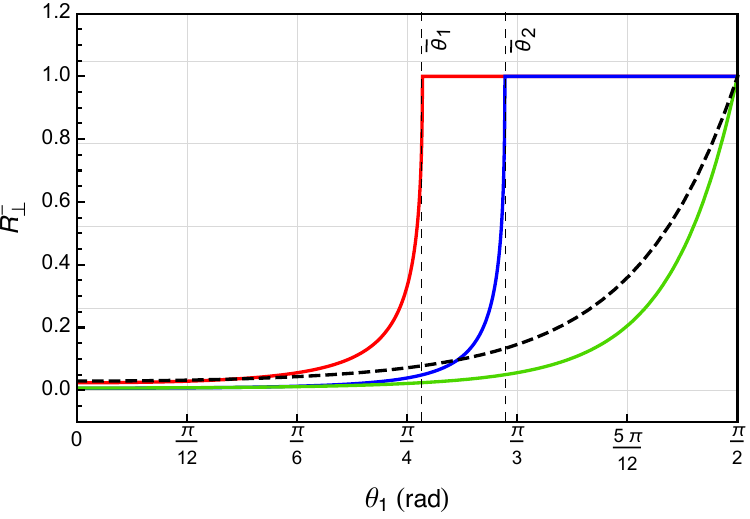}\caption{Reflection coefficient $R^{-}_{\perp}$ in terms of the incidence angle $\theta_{1}$. For the medium 1, we have used $\mu_{1}=1$, $\epsilon_{1}=1$, and for the medium 2: $\mu_{2}=1$, $\epsilon_{2}=2$, $\Sigma/\omega = 0.5$ (green line), $\Sigma/\omega = 1.5$ (blue line), and $\Sigma/\omega = 2$ (red) . The dashed black line represents the usual case ($\Sigma=0$). The vertical dashed lines indicate $\bar{\theta}_{1} \simeq 0.261\, \pi$ $\mathrm{rad}$ and $\bar{\theta}_{2} \simeq 0.324 \, \pi$ $\mathrm{rad}$, accordingly to \eqref{reflection-with-magnetic-conductivity-20}. }%
	\label{plot-R-perpendicular-menos-frequencia-fixa}%
\end{figure}

The behavior $R^{-}_{\perp}=1$ in the window $0<\omega  \leq \omega_{i}$ represents a scenario of total reflection induced by the non-null magnetic conductivity, $\Sigma$. Indeed, in such a region the refractive index $n_{2-}$ becomes less than $n_{1}$, so Snell's law yields total reflection. This low-frequency effect does not happen in the usual scenario of an ordinary dielectric - ordinary dielectric interface, whose non-dispersive behavior is represented by the horizontal lines in Fig.~\ref{plot-R-perpendicular-menos-angulos-fixos-with-USUAL-cases}. Besides this first peculiar aspect, the coefficient displays another interesting feature. It becomes null, $R^{-}_{\perp} =0$, for a specific value of frequency, $\omega_{0}$, given by
\begin{align}
\omega_{0} &= \frac{ \mu_{2} \, \Sigma \, \sqrt{ \mu_{1}\epsilon_{1}} \, \sqrt{ \mu_{1}^{2} \sin^{2} \theta_{1} + \mu_{2}^{2} \cos^{2} \theta_{1}} } {   \left|  \mu_{1} \mu_{2} \epsilon_{2} - \epsilon_{1} \left( \mu_{1}^{2} \sin^{2} \theta_{1} + \mu_{2}^{2} \cos^{2} \theta_{1} \right)  \right| } , \label{reflection-with-magnetic-conductivity-16} 
\end{align}
being also a consequence of non-null magnetic conductivity. For all examples in Fig.~\ref{plot-R-perpendicular-menos-angulos-fixos}, one finds $R^{-}_{\perp}=0$ for $\omega = 2$~$\mathrm{rad}$ $\mathrm{s}^{-1}$.

On the other hand, the general behavior of $R^{-}_{\perp}$ in terms of the incidence angle $\theta_{1}$ is displayed in Fig.~\ref{plot-R-perpendicular-menos-frequencia-fixa}, where one observes another peculiar property, given in terms of the regions fulfilling total reflection, that is, $R^{-}_{\perp}=1$. These regions occur for $\Sigma /\omega >1$ (low-frequency regime), being defined for $\bar{\theta}_{i} < \theta_{1} < \pi/2$, where 
\begin{align}
	\bar{\theta}_{i}^{\perp} &= \mathrm{arc} \sin \left[ \frac{1}{\sqrt{\mu_{1}\epsilon_{1}}} \left( \sqrt{  \mu_{2} \epsilon_{2} + \frac{\mu_{2}^{2}\Sigma^{2}}{4\omega^{2}} } - \frac{\mu_{2} \Sigma}{2\omega}  \right)  \, \right]. \label{reflection-with-magnetic-conductivity-20}
\end{align}

For $\Sigma/\omega <1$, the value of $\bar{\theta}_{i}$ becomes complex, meaning that the flat horizontal region, $R^{-}_{\perp}=1$, does not occur in the high-frequency limit or very low $\Sigma$, which includes the scenario of a conventional dielectric, $\Sigma=0$ (see the dashed black line in Fig.~\ref{plot-R-perpendicular-menos-frequencia-fixa}).

\subsection{Incident wave with $p$-polarization}

\begin{figure}[h]
\centering\includegraphics[scale=.69]{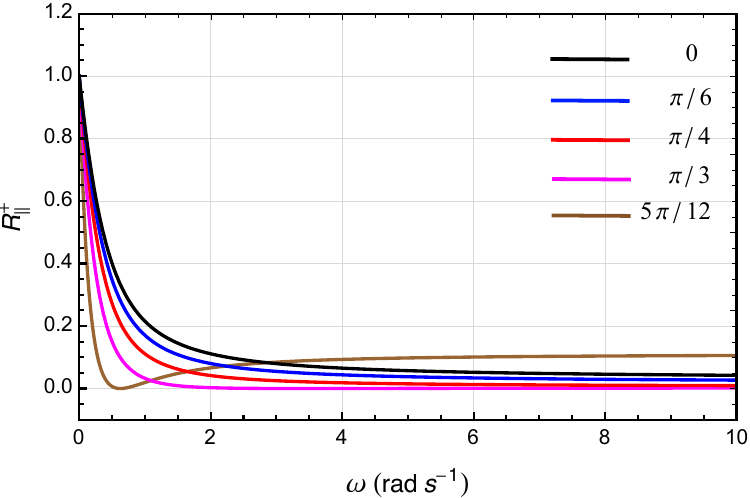}\caption{Reflection coefficient $R^{+}_{\parallel}$ in terms of frequency $\omega$ for fixed incidence angles. For the medium 1, we have used $\mu_{1}=1$, $\epsilon_{1}=1$, and for the medium 2: $\mu_{2}=1$, $\epsilon_{2}=2$, $\Sigma = 2$ $\mathrm{s}^{-1}$. }%
\label{plot-R-paralelo-mais-angulos-fixos}%
\end{figure}

\begin{figure}[h]
\centering\includegraphics[scale=.69]{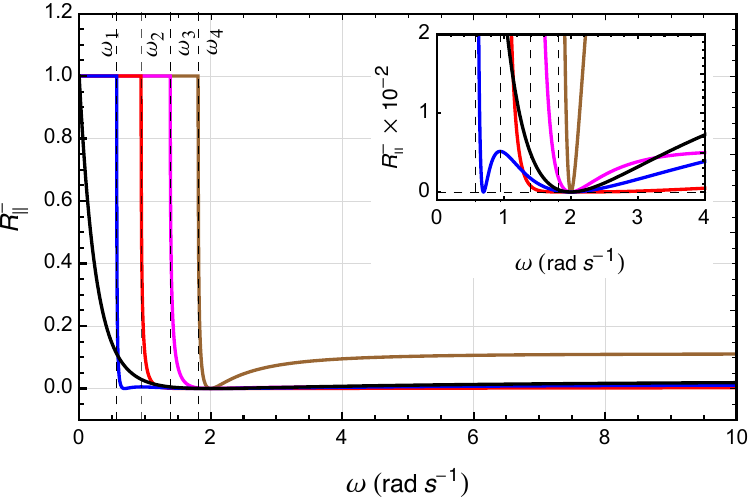}\caption{Reflection coefficient $R^{-}_{\parallel}$ in terms of frequency $\omega$ for fixed incidence angles (radians): $0$ (black), $\pi/6$ (blue), $\pi/4$ (red), $\pi/3$ (magenta), and $5\pi/12$ (brown). Here, we have used $\mu_{1}=\mu_{2}=1$, $\epsilon_{1}=1$, $\epsilon_{2}=2$ e $\Sigma = 2$ $\mathrm{s}^{-1}$. The inset plot highlights the behavior of $R_{\parallel}^{-}$ around its roots and shows that $R_{\parallel}^{-}$ has two roots in $\omega$ for $\theta_{1}=\pi/6$, differing from all the other $\theta_{1}$ cases, where only one root (in $\omega= 2$ $\mathrm{rad}$ $s^{-1}$) is found.}
\label{plot-R-paralelo-menos-angulos-fixos-modelo-2}
\end{figure}

For incident wave with $p$-polarization, the reflection coefficient is $R_{\parallel}^{\pm}$ of \eqref{reflection-with-magnetic-conductivity-9}.  Figures \ref{plot-R-paralelo-mais-angulos-fixos} and  \ref{plot-R-paralelo-menos-angulos-fixos-modelo-2} depict the reflectance $R_{\parallel}^{+}$ and $R_{\parallel}^{-}$, respectivelly,  in terms of the frequency. The profile of $R_{\parallel}^{+}$ may exhibit a frequency value for null reflection, which does not occur for $R_{\perp}^{+}$ (see Fig. \ref{plot-R-perpendicular-mais-angulos-fixos-with-USUAL-cases}).  Figure \ref{plot-R-paralelo-menos-angulos-fixos-modelo-2} should be compared with Fig. \ref{plot-R-perpendicular-menos-angulos-fixos}, revealing nearly the same qualitative behavior for $R^{-}_{\parallel}$ and $R^{-}_{\perp}$.

We point out that $R_{\parallel}^{\pm}=0$ can occur for specific frequency values, which are determined by the real and positive solutions of
\begin{align}
\omega \left(\mu_{1} \epsilon_{2}  -\mu_{2} \epsilon_{1} \, \frac{\cos^{2} \theta_{2\pm}}{\cos^{2}\theta_{1}} \right) \pm \mu_{2} \, \Sigma \sqrt{\mu_{1}\epsilon_{1}} \, \frac{\cos\theta_{2\pm}}{\cos\theta_{1}}  &=0, 
\end{align}
with $\theta_{2\pm}$ given by \eqref{reflection-with-magnetic-conductivity-12} and the upper and lower signs providing the solutions for $R^{+}_{\parallel}=0$ an $R^{-}_{\parallel}=0$, respectively.

\begin{figure}[h]
\centering\includegraphics[scale=.69]{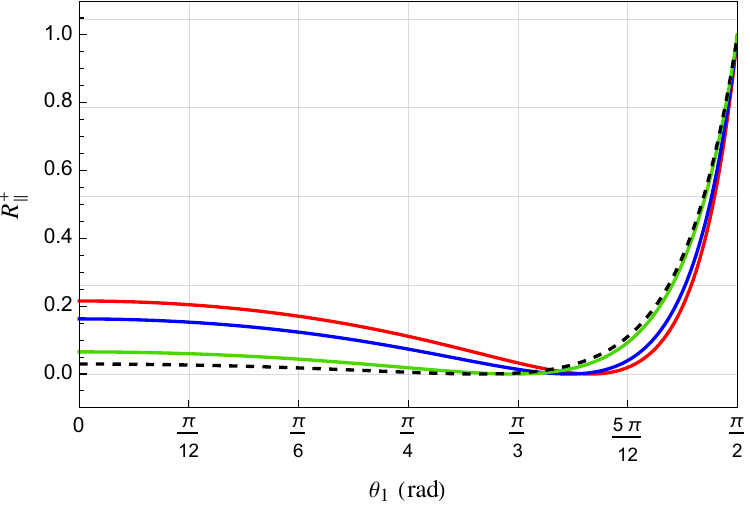}\caption{Reflection coefficient $R^{+}_{\parallel}$ in terms of the incidence angle $\theta_{1}$. Medium 1 defined by $\mu_{1}=1$, $\epsilon_{1}=1$, and medium 2 by $\mu_{2}=1$, $\epsilon_{2}=2$, $\Sigma/\omega = 0.5$ (green), $\Sigma/\omega = 1.5$ (blue), and $\Sigma/\omega = 2$ (red) . The dashed black line represents the usual case ($\Sigma=0$). }%
\label{plot-R-paralelo-mais-frequencia-fixa}%
\end{figure}

\begin{figure}[h]
\centering\includegraphics[scale=.69]{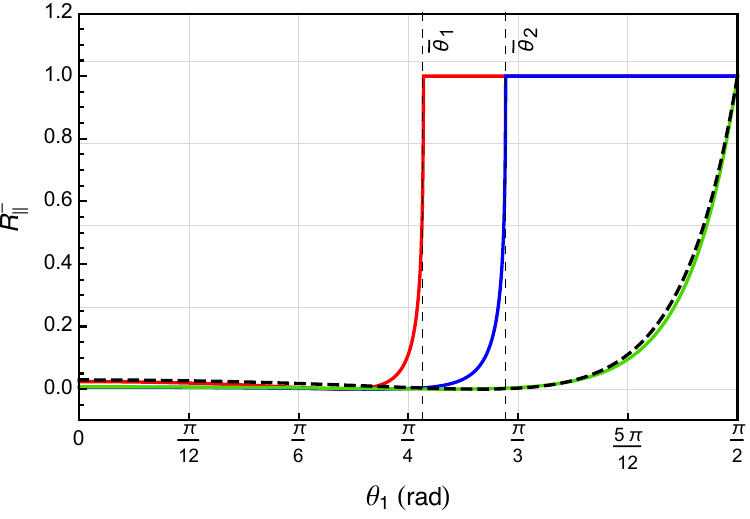}\caption{Reflection coefficient $R^{-}_{\parallel}$ in terms of the incidence angle $\theta_{1}$. For the medium 1, we have used $\mu_{1}=1$, $\epsilon_{1}=1$, and for the medium 2: $\mu_{2}=1$, $\epsilon_{2}=2$, $\Sigma/\omega = 0.5$ (black), $\Sigma/\omega = 1.5$ (blue), and $\Sigma/\omega = 2$ (red). The dashed black line represents the usual case ($\Sigma=0$). The vertical dashed lines indicate $\bar{\theta}_{1} \simeq 0.261\, \pi$ $\mathrm{rad}$ and $\bar{\theta}_{2} \simeq 0.324 \, \pi$ $\mathrm{rad}$, accordingly to \eqref{reflection-with-magnetic-conductivity-20-2A}. }%
\label{plot-R-paralelo-menos-frequencia-fixa}%
\end{figure}

The behavior of $R_{\parallel}^{\pm}$ with the incidence angle is illustrated in Figs.~\ref{plot-R-paralelo-mais-frequencia-fixa} and \ref{plot-R-paralelo-menos-frequencia-fixa}, respectively. Both are different from the coefficients $R^{+}_{\perp}$ of Fig. \ref{plot-R-perpendicular-mais-frequencia-fixa} and $R^{-}_{\perp}$ of Fig. \ref{plot-R-perpendicular-menos-frequencia-fixa}. As shown in Fig. \ref{plot-R-paralelo-mais-frequencia-fixa}, the coefficient $R_{\parallel}^{+}$ initially decreases with $\theta_{1}$, reaching a null value at the Brewster angle, and after increasing to the maximum value ($R_{\parallel}^{+}=1$) while the angle $\theta_{1}$ approaches $\pi/2$. This behavior differs from the $R^{+}_{\perp}$ profile, which grows up monotonically (see Fig. \ref{plot-R-perpendicular-mais-frequencia-fixa}). 

As for the behavior  of $R_{\parallel}^{-}$ with $\theta_{1}$, given in Fig. \ref{plot-R-paralelo-menos-frequencia-fixa}, there is a near similarity with the profile of $R^{-}_{\perp}$ of Fig. \ref{plot-R-perpendicular-menos-frequencia-fixa}, both presenting a window of total reflection starting in a given value of incidence angle,	$\bar{\theta}_{i}$, given by
	\begin{align}
		\bar{\theta}_{i}^{\parallel} &= \mathrm{arc} \sin \left[ \frac{1}{\sqrt{\mu_{1}\epsilon_{1}}} \left( \sqrt{  \mu_{2} \epsilon_{2} + \frac{\mu_{2}^{2}\Sigma^{2}}{4\omega^{2}} } - \frac{\mu_{2} \Sigma}{2\omega}  \right) \, \right]. \label{reflection-with-magnetic-conductivity-20-2A}
	\end{align}
The distinction between them rests in the slightly decreasing magnitude observed before reaching the Brewster angle.


\subsection{Total reflection}

In the last section, the total reflection was reported for $R_{\perp, \parallel}^{-}=1$ in the frequency window $0<\omega<\omega_{i}$, with $\omega_{i}$ of \eqref{reflection-with-magnetic-conductivity-14}. The total reflection also happens when the incidence angle is defined in the range $\bar{\theta}_{i} <\theta_{1} < \pi/2$, with $\bar{\theta}_{i}$ of \eqref{reflection-with-magnetic-conductivity-20} and \eqref{reflection-with-magnetic-conductivity-20-2A}, respectively, also written as
\begin{align}
	\bar{\theta}_{i}^{\pm} &= \mathrm{arc} \sin \left[ \frac{1}{\sqrt{\mu_{1}\epsilon_{1}}} \left( \sqrt{  \mu_{2} \epsilon_{2} + \frac{\mu_{2}^{2}\Sigma^{2}}{4\omega^{2}} } \pm \frac{\mu_{2} \Sigma}{2\omega}  \right) \, \right]. \label{reflection-with-magnetic-conductivity-20-2}
	\end{align}

Due to the values of the constitutive parameters adopted in the plots of the latter section, $\mu_{2} \epsilon_{2} > \mu_{1}\epsilon_{1}$, the total reflection was verified only for $R_{\parallel}^{-}$ and $R_{\perp}^{-}$. However, such a phenomenon can also occur for the polarization mode $+$ if we adopt the condition $\mu_{2} \epsilon_{2} < \mu_{1}\epsilon_{1}$, as shown in the first line of Table ~\ref{tab:conditions-for-total-reflection}. The situations of total reflection of Fig.~\ref{plot-R-paralelo-menos-angulos-fixos-modelo-2} are contained in the conditions of the second row and second column of Tab.~\ref{tab:conditions-for-total-reflection}.

 \begin{table}[h]
\caption{\small{Conditions for total reflection.}}
\centering
\begin{tabular}{ p{3.7cm}  p{4.3cm}        } 
\hline \hline 
 \\ [-2ex]
\centering\arraybackslash$R_{\perp, \parallel}^{+} =1$ & \centering\arraybackslash $R_{\perp, \parallel}^{-} =1$   \\ [1ex]
\hline
\\ [-1ex]
\hspace{.5cm} $\mu_{2} \epsilon_{2} < \mu_{1} \epsilon_{1} $, \newline\newline \centering\arraybackslash $\omega >  \frac{ \displaystyle \mu_{2} \, \sin\theta_{1}  \,\Sigma \sqrt{\mu_{1}\epsilon_{1}}} {\displaystyle \left| \mu_{1}\epsilon_{1} \sin^{2} \theta_{1} - \mu_{2} \epsilon_{2} \right|} .$ &    \hspace{.5cm}  $\mu_{2} \epsilon_{2} \leq \mu_{1} \epsilon_{1}  $, \newline  \newline   \centering\arraybackslash    $\omega >0 .$           \\ [3ex]   
\hline \\
     \centering\arraybackslash  \vspace{0.1cm}  $\mu_{2} \epsilon_{2} > \mu_{1}\epsilon_{1},$ \newline no range in $\omega$ for it. & $\mu_{2} \epsilon_{2} > \mu_{1}\epsilon_{1} ,$ \newline \newline \centering\arraybackslash $0 < \omega \leq  \frac{ \displaystyle \mu_{2} \, \sin\theta_{1} \, \Sigma \sqrt{\mu_{1}\epsilon_{1}  }} {\displaystyle \left| \mu_{2}\epsilon_{2} -  \mu_{1}\epsilon_{1} \sin^{2}\theta_{1} \right|  }   .$ \\[3ex]
     \hline \\
 $\mu_{1}=\mu_{2}, \, \epsilon_{1}=\epsilon_{2},$ \newline 
 \centering\arraybackslash $\Sigma <0 , \, \omega>0. $     &   \centering\arraybackslash   $\mu_{1}=\mu_{2}, \, \epsilon_{1}=\epsilon_{2},$ \newline no range in $\omega$ for it.
 \\[3ex]
\hline \hline
\end{tabular}
\label{tab:conditions-for-total-reflection}
\end{table}


Table 1 also contains a special and interesting possibility of total reflection for $\epsilon_{1}=\epsilon_{2}$ and $\mu_{1}=\mu_{2}$. But in this case, the total reflection can only occur when $\Sigma<0$ for $R_{\parallel, \perp}^{+}=1$ (for any value of frequency), while $R_{\perp, \parallel}^{-}=1$ does not happen under these conditions. This theoretical possibility will not be discussed in this work, but it may motivate additional investigations in the future.

\section{\label{Critical-angles-section}Critical angles for null reflection and polarization change}

In this section, we analyze the critical angles that determine the incidence configuration at which the reflection becomes null. Since we have obtained two reflection coefficients for each scenario regarding the polarizations ($s$ or $p$) of the incident wave, there will be two distinct critical angles, one for each mode.

\subsection{Critical angles for incident $s$-polarization wave}

Starting from the coefficient  (\ref{reflection-with-magnetic-conductivity-6}), the condition for a real non-null critical angles $\theta_{1\pm}^{c}$, for which $R^{\pm}_{\perp} =0$, is given by
	\begin{align}
	\tan^{2} (\theta_{s}^{c})_\pm &= \frac{ \mu_{2}^{2} n_{1}^{2}/  \mu_{1}^{2} - n_{2\pm}^{2} } {\left( n_{2\pm}^{2} -n_{1}^{2} \right) } , \label{reflection-with-magnetic-conductivity-21A}
\end{align}
which also is expressed as
\begin{align}
\tan^{2} \theta^{c}_{s\pm} &=  \frac{ \displaystyle    \frac{\mu_{2}^2 \epsilon_{1}}{\mu_{1}} - \left(   \sqrt{\mu_{2} \epsilon_{2} + \left( \frac{ \mu_{2} \Sigma}{2\omega} \right)^{2}} \pm \frac{\mu_{2} \Sigma}{2\omega} \right)^{2}  }  {    \displaystyle        \left(  \sqrt{\mu_{2} \epsilon_{2} + \left( \frac{ \mu_{2} \Sigma}{2\omega} \right)^{2}} \pm \frac{\mu_{2} \Sigma}{2\omega} \right)^{2} - \mu_{1} \epsilon_{1} }  . \label{reflection-with-magnetic-conductivity-21}
\end{align}
In general, these angles
	 constitute partial Brewster angles, since they state the incidence condition that turns null one of the two $s$-modes. However, if exists one frequency value for which both angles become equal, such a frequency defines a total Brewster angle for $s$ polarization).

\begin{figure}[h]
\centering\includegraphics[scale=.69]{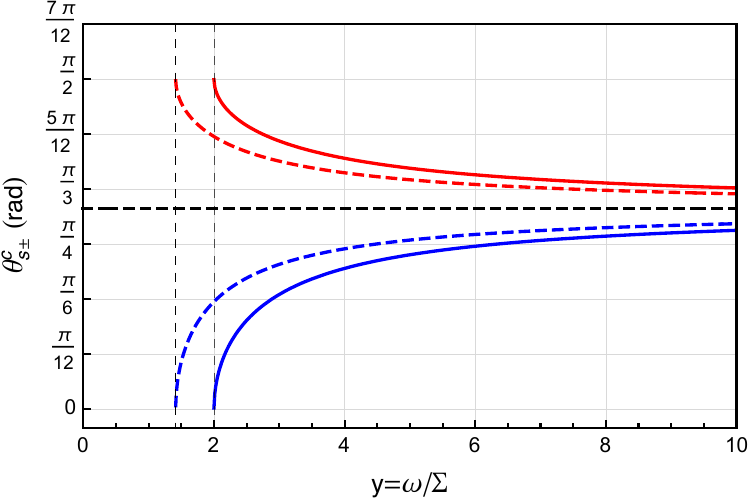}\caption{Critical angles $\theta_{1+}^{c}$ (azul) and $\theta_{1-}^{c}$  (red)  in terms of frequency $y=\omega/\Sigma$. The solid curves indicate the critical angles when the medium 1 is described by $\mu_{1}=1$, $\epsilon_{1}=1$, and the medium 2 by $\mu_{2}=2$, $\epsilon_{2}=1$. For the color-dashed lines, one has $\mu_{1}=1$, $\epsilon_{1}=2$ and $\mu_{2}=2$, $\epsilon_{2}=2$.  The dashed black horizontal line indicates the usual case ($\Sigma=0$). The dashed vertical lines define the lower limit for $y$, given by \eqref{reflection-with-magnetic-conductivity-22}.}
\label{plot-critical-angles-R-perpendicular-exemplo-1}
\end{figure}

Considering that the constitutive parameters ($\epsilon_{1,2}, \mu_{1,2}, \Sigma$) are real and positive, the general conditions to determine the critical angles are given in Tables \ref{tab:conditions-for-critical-angles-R-perpendicular-1} and \ref{tab:conditions-for-critical-angles-R-perpendicular-2} of Appendix \ref{Tables-conditions-for-null-reflection}. We observe in $c)$ row of these tables that $R_{\perp}^{+} =0$ and $R_{\perp}^{-}=0$ can occur simultaneously for the same set of conditions\footnote{If one chooses another set condition from Tab.~\ref{tab:conditions-for-critical-angles-R-paralelo-mais} (or \ref{tab:conditions-for-critical-angles-R-paralelo-menos}) of Appendix \ref{Tables-conditions-for-null-reflection}, for instance, one would obtain only a critical angle for $R_{\parallel}^{+}=0$ (or for $R_{\parallel}^{-}=0$), since the existence of only one critical angle would be granted. This explains the reason to have chosen the sets of row $c)$ of Tabs. \ref{tab:conditions-for-critical-angles-R-perpendicular-1} and \ref{tab:conditions-for-critical-angles-R-perpendicular-2}, which are able to provide both $R_{\perp}^{+}=0$ and $R_{\perp}^{-}=0$.}. However, the corresponding critical angles, $\theta_{s+}^{c}$ and $\theta_{s-}^{c}$, are distinct, since they are defined by different expressions, see the $\pm$ sign in \eqref{reflection-with-magnetic-conductivity-21}. Consequently, one has a specific critical angle at which $R_{\perp}^{+} =0$, $R_{\perp}^{-} \neq 0$ and another angle for $R_{\perp}^{-}=0$, $R_{\perp}^{+} \neq 0$. To illustrate the behavior of the critical angles, Figure \ref{plot-critical-angles-R-perpendicular-exemplo-1} shows the critical angles, $\theta_{s+}^{c}$ and $\theta_{s-}^{c}$, in terms of the dimensionless parameter $y=\omega / \Sigma$, for the conditions given in row $d)$ of tables \ref{tab:conditions-for-critical-angles-R-perpendicular-1} and \ref{tab:conditions-for-critical-angles-R-perpendicular-2}.

In Fig.~\ref{plot-critical-angles-R-perpendicular-exemplo-1}, the dashed vertical lines indicate the lower bound for frequency $\omega$, i. e., they stipulate the window $y>\hat{y}$ in which the critical angles can exist, with $\hat{y}$ given by
\begin{align}
\hat{y}= \sqrt{ \frac{\mu_{1}}{\epsilon_{1}}} \,  \frac{\mu_{2} }{   \mu_{2} - \mu_{1}  }, \label{reflection-with-magnetic-conductivity-22}
\end{align}
and $\mu_{2} >\mu_{1}$, since we are considering the conditions of row $c)$ of Tabs. \ref{tab:conditions-for-critical-angles-R-perpendicular-1} and \ref{tab:conditions-for-critical-angles-R-perpendicular-2}. As the corresponding curves for $\theta_{s+}^{c}$ (azul) and $\theta_{s-}^{c}$ (red) in Fig.~\ref{plot-critical-angles-R-perpendicular-exemplo-1} never cross each other, there is no frequency in which $\theta_{1+}^{c} = \theta_{1-}^{c}$. The absence of a unique angle yielding simultaneously $R_{\perp}^{+}=0$ and $R_{\perp}^{-}=0$ avoids the definition of a total Brewster angle (for total null reflection for the $s$-polarization). Therefore, for the set of examined parameters, only partial $s$-Brewster angles were reported.

\subsection{Circular Polarized reflected wave for $s$-polarization incident wave}

The existence of partial Brewster angles engenders the polarization change of the reflected wave in relation to incident one. Indeed, given a linear s-polarized wave, which can be regarded as the sum of RCP and LCP waves, $E_{0s}=E_{-}+E_{+}$, respectively, the reflected ray can emerge as circularly polarized, as follows: 
 \begin{itemize} 
 \item For the case $\theta=\theta_{s+}^{c}$, the LCP wave is suppressed and the reflected wave contains only the piece $E_{-}$, becoming RCP.
 \item For the case $\theta=\theta_{s-}^{c}$, the RCP wave is suppressed and the reflected wave contains only the piece $E_{+}$, becoming LCP.
\end{itemize}

For the special case of non-magnetic materials ($\mu_{1}=\mu_{2} \simeq \mu_{0}$), expression (\ref{reflection-with-magnetic-conductivity-21A}) yields $\tan^{2} \theta_{1\pm}^{c} = -1$, which has no real solutions. Therefore, there are no critical angles for $s$-polarized incident wave at the interface between non-magnetic media, as it occurs in the usual dielectric reflection. This is due to the fact the expression (\ref{reflection-with-magnetic-conductivity-21A}) has the same structure as the usual dielectric-dielectric interface. 
 The next section shows that this special case can occur for the incident wave $p$-polarized.

\subsection{Critical angles and polarization change for $p$-polarization incident wave}

When the incident wave is $p$-polarized, the coefficient  (\ref{reflection-with-magnetic-conductivity-7}) yields critical angles defined by
\begin{align}
\tan^{2} \theta_{p\pm}^{c} &= \frac{ \mu_{1}^{2} n_{2\pm}^{4} - \mu_{2}^{2} n_{1}^{2} n_{2\pm}^{2} } {\mu_{2}^{2} n_{1}^{2} \left( n_{2\pm}^{2} -n_{1}^{2} \right) } , \label{reflection-with-magnetic-conductivity-23}
\end{align}
with $n_{1}^{2} = \mu_{1}\epsilon_{1}$ and $n_{2\pm}$ given in \eqref{reflection-with-magnetic-conductivity-5}. The angles (\ref{reflection-with-magnetic-conductivity-23}) only occur when the constitutive parameters fulfill one of the several possible sets of conditions in Tabs. \ref{tab:conditions-for-critical-angles-R-paralelo-mais} and \ref{tab:conditions-for-critical-angles-R-paralelo-menos} for real and positive constitutive parameters. 

\begin{figure}[h]
	\centering\includegraphics[scale=.69]{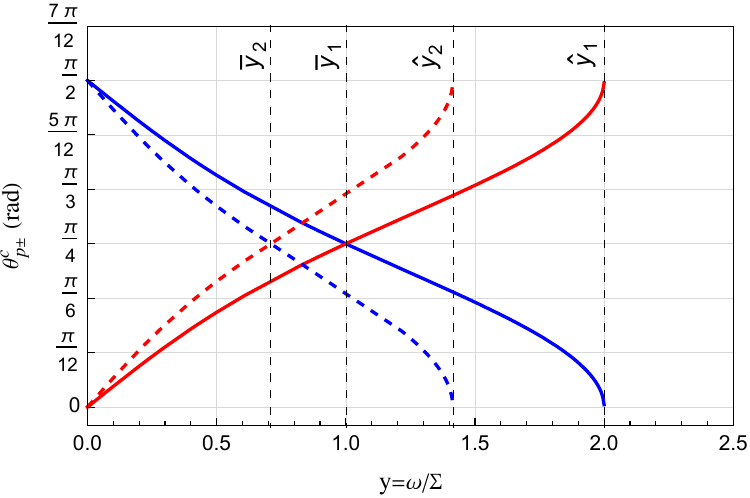}\caption{Critical angles $\theta_{1+}^{c}$ (azul) and $\theta_{1-}^{c}$  (red)  in terms of frequency $y=\omega/\Sigma$. Here, we have used $\mu_{1}=1$, $\mu_{2}=2$, $\epsilon_{1}=\epsilon_{2}=1$ (solid curves), and $\epsilon_{1}=\epsilon_{2}=2$ (dashed lines).  The dashed vertical lines at $\hat{y}_{1,2}$ define the upper limit for $y$, according to \eqref{reflection-with-magnetic-conductivity-22}.  The dashed vertical lines at $\bar{y}_{1,2}$, given by \eqref{reflection-with-magnetic-conductivity-24}, define the values of $y$ where $\theta_{1+}^{c}=\theta_{1-}^{c}$. Notice that the usual scenario ($\Sigma=0$) is not depicted since \eqref{reflection-with-magnetic-conductivity-23} has no real solutions when $\mu_{2} >\mu_{1}$, $\epsilon_{2}=\epsilon_{1}$ and $\Sigma=0$. }
	\label{plot-critical-angles-R-paralelo-exemplo-1}%
\end{figure}

Figure \ref{plot-critical-angles-R-paralelo-exemplo-1} depicts the critical angles $\theta_{p\pm}^{c}$ in terms of the dimensionless parameter $y=\omega/\Sigma$ for the conditions given in the row $c)$ of Tab. \ref{tab:conditions-for-critical-angles-R-paralelo-mais} and row $d)$ of Tab. \ref{tab:conditions-for-critical-angles-R-paralelo-menos}, whose choice allows us to represent $\theta_{p+}^{c}$ and $\theta_{p-}^{c}$ in the same graph, due to the same reason explained in the latter ``s"-polarization section.

According to the set of conditions adopted, both the critical angles are only defined when $y<\tilde{y}$, with $\tilde{y}$ given in \eqref{reflection-with-magnetic-conductivity-24}. Differently from the previous $s$-polarized scenario, the present case allows $\theta_{p+}^{c} =\theta_{p-}^{c}$ for a certain value of $y$, designated as $\bar{y}=\bar{\omega}/ \Sigma$, and given by
\begin{align}
	\bar{y}&= \sqrt{ \frac{\mu_{1}}{\epsilon_{1}}}  \, \sqrt{ \frac{\mu_{2}}{2 \left(\mu_{2} - \mu_{1} \right)}}  \label{reflection-with-magnetic-conductivity-24} .
\end{align}
Thus, by using $\bar{y}$, one determines the frequency $\bar{\omega}$ for which one simultaneously has $R_{\parallel}^{+}=0$ and $R_{\parallel}^{-}=0$, defining a total Brewster angle, labeled as $\left. \theta_{p}\right|_{\bar{\omega}}$, at which it also occurs polarization modification. Indeed, for the particular conditions adopted in rows $c)$ of Tab. \ref{tab:conditions-for-critical-angles-R-paralelo-mais} and row $d)$ of Tab. \ref{tab:conditions-for-critical-angles-R-paralelo-menos}, we observe a total Brewster angle of $\pi/4$. Furthermore, considering a mixed incident linearly polarization written as $E_{0}=E_{s}+E_{p}$, the reflected ray becomes $s$-polarized, $E_{R}=E_{s}$. For other values of frequency, $\omega \neq \bar{\omega}$, in which only partial Brewster angles hold, there occurs a change of $p$-linear polarization, $E_{0p}=E_{-}+E_{+}$, to the circular one, as follows: 
	\begin{itemize} 
		\item For the case $\theta=\theta_{p+}^{c}$, the LCP wave is suppressed and the reflected wave contains only the piece $E_{-}$, becoming RCP.
		\item For the case $\theta=\theta_{p-}^{c}$, the RCP wave is suppressed and the reflected wave contains only the piece $E_{+}$, becoming LCP.
\end{itemize}

The fact that the expression (\ref{reflection-with-magnetic-conductivity-21A}) has the same structure of the usual dielectric-dielectric interface implies equal results for the $p$-polarization, that is, 
\begin{itemize}
	\item For $\mu_{1}=\mu_{2}$, one has $\tan \theta_{p\pm}^{c}=n_2/n_1$, which implies
	\begin{equation}
		\theta_{p\pm}^{c}+\theta_{2}=\pi/2.
	\end{equation}
	\item For $\mu_{1}>\mu_{2}$, one has $\tan \theta_{p\pm}^{c}>n_2/n_1$, which implies
	\begin{equation}
	\theta_{p\pm}^{c}+\theta_{2}<\pi/2.
	\end{equation}
	\item For $\mu_{1}<\mu_{2}$ (and $\mu_{1}n_{2\pm}>\mu_{2}n_1$), one has $\tan \theta_{p\pm}^{c}<n_2/n_1$, which yields
	\begin{equation}
		\theta_{p\pm}^{c}+\theta_{2}>\pi/2.
	\end{equation}
\end{itemize}

\section{\label{other-optical-effects-section}Optical effects}

In this section, we discuss other relevant optical properties of the reflected wave, namely the Goos-H\"anchen effect  (GH) and the complex Kerr rotation angle in the reflection.

\subsection{Goos-H\"anchen effect}

When the incidence angle is greater than 
\begin{align}
\bar{\theta}_{i}^{\pm} &= \mathrm{arc} \sin \left[ \frac{1}{\sqrt{\mu_{1}\epsilon_{1}}} \left( \sqrt{  \mu_{2} \epsilon_{2} + \frac{\mu_{2}^{2}\Sigma^{2}}{4\omega^{2}} } \pm \frac{\mu_{2} \Sigma}{2\omega}  \right)  \, \right], \label{reflection-with-magnetic-conductivity-25}
\end{align}
total reflection occurs for the corresponding propagating mode. In this scenario, the associated reflected wave undergoes a lateral shift, a displacement of the reflected wave relative to the position in which there is no total reflection, given as
\begin{align}
D&=- \frac{\lambda}{2\pi} \,  \frac{ \partial \varphi }{\partial \theta_{1}}, \label{reflection-with-magnetic-conductivity-26}
\end{align}
where $\lambda$ is the wavelength of the incident light and $\varphi$ is the phase acquired during total internal reflection\footnote{In a standard reflection scenario between two usual dielectrics ($\Sigma=0$), to enable the condition for total reflection and the displacement occurrence, the incident wave travels from medium 1 to medium 2, with $n_1 >n_2$.}. 
Such a lateral optical displacement of the reflected ray is known as the Goos-H\"anchen effect \cite{Jackson, Ming-Che}.

\subsubsection{Incident wave with $s$-polarization}

Considering the $s$-polarized incident wave, the reflection coefficient $r_{\perp}^{\pm}$ of \eqref{reflection-with-magnetic-conductivity-3}, in the regime of total internal reflection, can be rewritten as $r_{\perp}^{\pm}= \exp \left(i \varphi_{s}^{\pm}\right) $, with the phase factor
\begin{align}
\varphi_{s}^{\pm} &= -2 \,  \mathrm{arc} \tan \left( \frac{\mu_{1}}{\mu_{2}} \, \frac{ \sqrt{ \sin^{2} \theta_{1} - \sin^{2}\bar{\theta}_{i}^{\pm}}}{\cos \theta_{1}} \right) . \label{reflection-with-magnetic-conductivity-27}
\end{align}
The GH displacement is obtained by replacing the phase (\ref{reflection-with-magnetic-conductivity-27}) in \eqref{reflection-with-magnetic-conductivity-26}, providing
\begin{subequations}
\label{reflection-with-magnetic-conductivity-27-a}
\begin{align}
D_{\perp}^{\pm} &=  \frac{ \lambda}{\pi} \frac{\sin \theta_{1}}{\sqrt{\sin^{2}\theta_{1}-\sin^{2}\bar{\theta}_{i}^{\pm}} }   \, g_{\pm} , \label{reflection-with-magnetic-conductivity-27-b} \\
g_{\pm}&= \frac{ \mu_{1}\mu_{2}  \cos^{2} \bar{\theta}_{i}^{\pm}}{ \mu_{2}^{2} \cos^{2}\theta_{1} + \mu_{1}^{2} \left(\sin^{2}\theta_{1} - \sin^{2} \bar{\theta}_{i}^{\pm} \right)} . \label{reflection-with-magnetic-conductivity-27-c}
\end{align} 
\end{subequations}

\begin{figure}[h]
\centering\includegraphics[scale=.69]{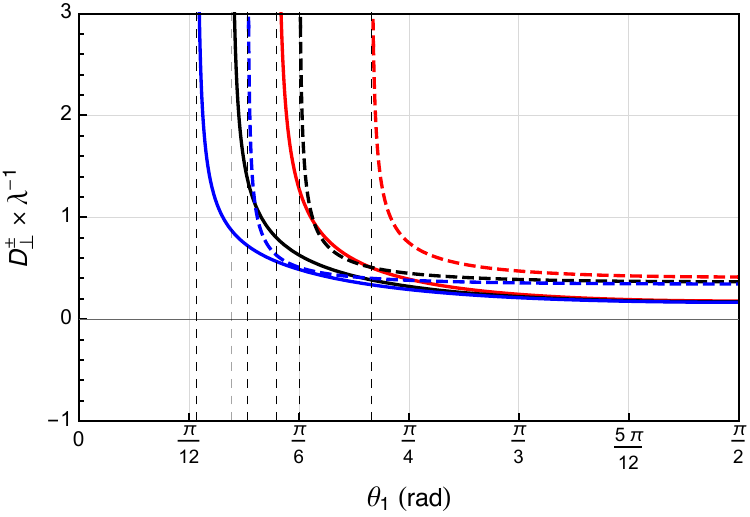}\caption{The Goos-H\"anchen shift $D_{\perp}^{+}$ (red lines) and $D_{\perp}^{-}$ (blue lines) in terms of the incidence angle, $\theta_1$. The dashed lines were obtained for $\epsilon_{1}=4$, $\epsilon_{2}=1$, $\Sigma/\omega = 0.5$, $\mu_{1}=1$, $\mu_{2}=1$ and represent the GH shift for a reflection in non-magnetic media. The solid lines correspond to $\epsilon_{1}=4$, $\epsilon_{2}=1$, $\Sigma/\omega = 0.5$ and $\mu_{1}=2, \mu_{2}=1$, representing the GH shift for a reflection with a magnetic medium. The black lines indicate the usual case ($\Sigma=0$). The vertical pale dashed lines indicate the critical angle above which total internal reflection occurs for each example, accordingly to \eqref{reflection-with-magnetic-conductivity-25}. }
\label{plot-Goos-Hanchen-s-polarization}%
\end{figure}

In figure \ref{plot-Goos-Hanchen-s-polarization}, we depict the lateral shift per unit length in terms of the incidence angle for magnetic and non-magnetic cases. Notice that the values of the constitutive parameters chosen here satisfy the general conditions given in the first row of the two columns of Tab.~\ref{tab:conditions-for-total-reflection}. This explains why one can represent in the same plot both $D^{\pm}_{\perp}$.

For all the cases examined, the GH shift begins large for angles immediately greater than the total reflection value and decreases monotonically until reaching an asymptotical low value when the incidence angle approaches $\pi/2$. For the case the interface separates non-magnetic media, the GH asymptotic shift may be a bit larger ($D_{\perp}^{+}$) or smaller ($D_{\perp}^{-}$) than the one obtained in the usual case ($\Sigma=0$) - see the black dashed line in Fig. \ref{plot-Goos-Hanchen-s-polarization}. On the other hand, for the case where the interface separates magnetic media ($\mu > 1$), the GH asymptotic shift is nearly equal ($D_{\perp}^{+}$ and $D_{\perp}^{-}$) to the one obtained in the usual case ($\Sigma=0$).
	
	So, we notice that the magnetic conductivity may alter the GH shift in reflections in non-magnetic media.

\subsubsection{Incident wave with $p$-polarization}

For incident wave $p$-polarized, the complex reflection coefficient is now $r_{\parallel}^{\pm}= exp \left( i \varphi_{p}^{\pm} \right)$ where
\begin{align}
\varphi_{p}^{\pm} &= -2 \, \mathrm{arc} \tan \left( \frac{\mu_{2}}{\mu_{1}} \frac{ \sqrt{ \sin^{2} \theta_{1} - \sin^{2} \bar{\theta}_{i}^{\pm} }} {\sin^{2} \bar{\theta}_{i}^{\pm} \, \cos \theta_{1} }\right)  . \label{reflection-with-magnetic-conductivity-28}
\end{align}
In this case, the GH shift is given by
\begin{subequations}
\label{reflection-with-magnetic-conductivity-29-0}
\begin{align}
D_{p}^{\pm} &= \frac{\lambda}{\pi} \frac{ \sin \theta_{1}}{ \sqrt{ \sin^{2} \theta_{1} - \sin^{2} \bar{\theta}_{i}^{\pm} }} \, h_{\pm} , \label{reflection-with-magnetic-conductivity-29-1} \\
h_{\pm} &=  \frac{ \mu_{1}\mu_{2} \sin^{2}\bar{\theta}_{i}^{\pm} \cos^{2} \bar{\theta}_{i}^{\pm}} { \mu_{1}^{2} \sin^{4} \bar{\theta}_{i}^{\pm} \cos^{2} \theta_{1} + \mu_{2}^{2} \left( \sin^{2} \theta_{1} - \sin^{2} \bar{\theta}_{i}^{\pm} \right) } . \label{reflection-with-magnetic-conductivity-29-2}
\end{align}
\end{subequations}

\begin{figure}[h]
\centering\includegraphics[scale=.69]{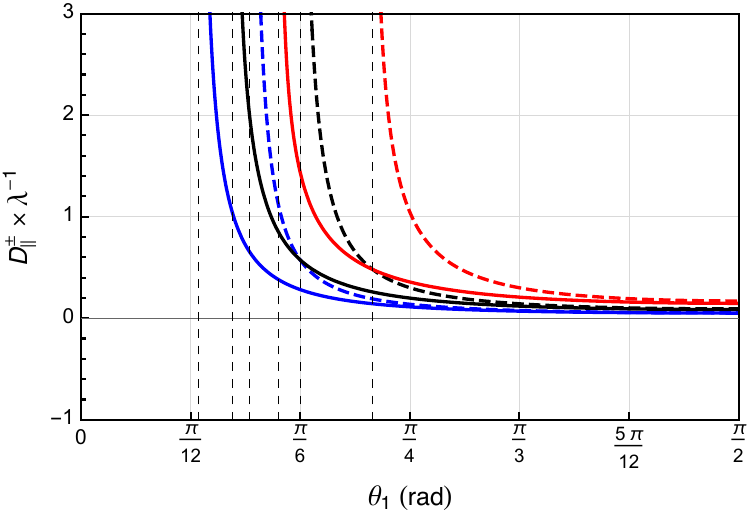}\caption{Goos-H\"anchen shift $D_{\parallel}^{+}$ (red lines) and $D_{\parallel}^{-}$ (blue lines). Here, we have used: $\epsilon_{1}=4$, $\epsilon_{2}=1$, $\Sigma/\omega = 0.5$, $\mu_{1}=\mu_{2}=1$ (dashed curves), and $\mu_{1}=2, \mu_{2}=1$ for solid lines. The black lines indicate the usual case ($\Sigma=0$). The vertical dashed lines indicate the critical angle above which total internal reflection occurs for each example, accordingly to \eqref{reflection-with-magnetic-conductivity-25}. }
\label{plot-Goos-Hanchen-p-polarization}%
\end{figure}

The GH displacement per unit length in terms of the incidence angle is represented in Fig.~\ref{plot-Goos-Hanchen-p-polarization}.

In general, Figures  \ref{plot-Goos-Hanchen-s-polarization} and \ref{plot-Goos-Hanchen-p-polarization} indicate that the GH displacement for each propagating mode is shifted from the standard cases (black curves). On one hand, we observe that the magnetic conductivity $\Sigma$ enables angles $\bar{\theta}^{\parallel, \perp}_{-}$ for total reflection smaller than the usual scenario, yielding, blue lines on the left of the black curves (usual case) in Figs.~\ref{plot-Goos-Hanchen-p-polarization} and \ref{plot-Goos-Hanchen-s-polarization}. On the other hand, the conductivity $\Sigma$ allows angles $\bar{\theta}^{\parallel, \perp}_{+}$ for total reflection greater than the usual scenario, yielding the red lines on the right of the black curves (usual case) in Figs.~\ref{plot-Goos-Hanchen-p-polarization} and \ref{plot-Goos-Hanchen-s-polarization}.

\subsection{Kerr rotation and Kerr ellipticity}

As demonstrated in \eqref{eq:polarizations-isotropic}, the propagating modes in the dielectric substrate with magnetic current are given by LCP and RCP vectors. In this way, for a reflected wave in the surface of such a medium,  the complex Kerr angle \cite{Shinagawa, Sato, Argyres} is defined by 
\begin{align}
\tan \Phi_{K} =  i \left( \frac{ r_{+} - r_{-}} {r_{+} + r_{-} } \right), \label{reflection-with-magnetic-conductivity-30}
\end{align}
where $r_{\pm}$ are the complex Fresnel coefficients for LCP and RCP polarizations for normal incidence. Indeed, by considering normal incidence, $\theta_{1}=0$, one finds\footnote{We are considering the sign convention of Ref.~\cite{Jackson, Hecht}, where, at normal incidence, the reflection coefficient is that one obtained for an incident wave with $p$-polarization. This happens because, as pointed out in Refs.~\cite{Zangwill}, at normal incidence there is no physical difference between $s$- and $p$- polarizations.}
\begin{align}
r_{\pm} &=\frac{\mu_{1} n_{2\pm} -  \mu_{2}  n_{1}}{ \mu_{1} n_{2\pm}  + \mu_{2} n_{1} } . \label{reflection-with-magnetic-conductivity-31} 
\end{align}

Since the reflection coefficients are, in general, complex, the Kerr rotation angle ($\theta_{K}$) and the Kerr ellipticity angle ($\eta_{K}$) are given by
\begin{align}
 \tan (2\theta_{K})&= - \frac{ 2 \Delta^{\prime\prime}  }{1 - |\Delta|^{2}}, \label{reflection-with-magnetic-conductivity-33} \\
\sin(2 \eta_{K})&=  \frac{2 \Delta^{\prime} }{1 + |\Delta|^{2}}, \label{reflection-with-magnetic-conductivity-34}
\end{align}
with $\Delta=\Delta^{\prime} + i \Delta^{\prime\prime}$, where $\Delta^{\prime} =\mathrm{Re}[ \Delta]$, $\Delta^{\prime\prime} = \mathrm{Im}[\Delta]$ and
\begin{align}
\Delta &= \frac{r_{+} - r_{-} } {r_{+} + r_{-} } . \label{reflection-with-magnetic-conductivity-34-1}
\end{align}

In the small angle approximation \cite{Shinagawa, Sato}, one can also write 
	\begin{align}
		\theta_{K}&= - \Delta^{\prime \prime}=  -\mathrm{Im} \left( \frac{ r_{+} - r_{-}} {r_{+} + r_{-} } \right),\label{reflection-with-magnetic-conductivity-33-1} \\
		\eta_{K}&=\Delta^{\prime} = \mathrm{Re} \left( \frac{ r_{+} - r_{-}} {r_{+} + r_{-} } \right), \label{reflection-with-magnetic-conductivity-34-2}
	\end{align}
	which happens for most usual materials \cite{Schlenker-Souche}. Implementing \eqref{reflection-with-magnetic-conductivity-31} in \eqref{reflection-with-magnetic-conductivity-34-1}, it provides
\begin{align}
\Delta &=  \frac{ \mu_{1} \mu_{2} n_{1} \left(n_{2+}-n_{2-} \right)} {\mu_{1}^{2} n_{2+} n_{2-} - \mu_{2}^{2} n_{1}^{2}} . \label{reflection-with-magnetic-conductivity-35} 
\end{align}
Replacing $n_{1} =\sqrt{\mu_{1} \epsilon_{1}}$ and \eqref{reflection-with-magnetic-conductivity-5} in \eqref{reflection-with-magnetic-conductivity-35}, one finds
\begin{align}
\Delta &= \frac{ \sqrt{\mu_{1} \epsilon_{1}} }{ \mu_{1}\epsilon_{2} - \mu_{2} \epsilon_{1} } \frac{\mu_{2} \Sigma}{\omega}.  \label{reflection-with-magnetic-conductivity-36}
\end{align}
Finally, by applying the real result (\ref{reflection-with-magnetic-conductivity-36}) in \eqref{reflection-with-magnetic-conductivity-33-1} and \eqref{reflection-with-magnetic-conductivity-34-2}, we obtain $\theta_{K} = 0$ and
\begin{align}
\sin (2 \eta_{K}) &=  \frac{ 2 \sqrt{\mu_{1} \epsilon_{1}} }{ \mu_{1}\epsilon_{2} - \mu_{2} \epsilon_{1} } \frac{\mu_{2} \Sigma}{\omega} \, \frac{1}{1+ \frac{\mu_{1}\epsilon_{1}}{(\mu_{1}\epsilon_{2} - \mu_{2} \epsilon_{1} )^{2}} \left( \frac{\mu_{2} \Sigma}{\omega} \right)^{2}} . \label{reflection-with-magnetic-conductivity-37-0}
\end{align}

\begin{figure}[h]
\centering\includegraphics[scale=.69]{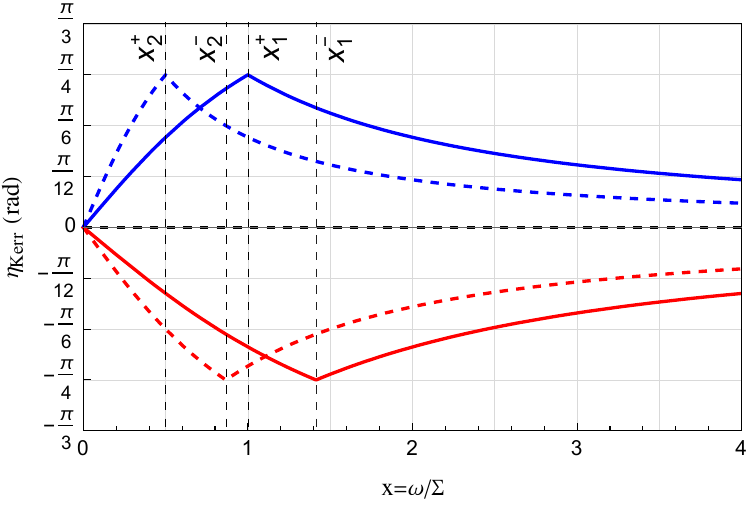}\caption{Kerr ellipticity angle of \eqref{reflection-with-magnetic-conductivity-37-0}. Here, we have used $\mu_{1}=\mu_{2}=1$ for all curves. For solid blue (red) lines, we used: $\epsilon_{1} =1$, $\epsilon_{2}=2$ ($\epsilon_{1}=2$, $\epsilon_{2}=1$). For dashed blue (red) curves, we used: $\epsilon_{1}=1$, $\epsilon_{2}=3$ ($\epsilon_{1}=3$, $\epsilon_{2}=1$). The horizontal dashed black line represents the usual case ($\Sigma=0$). The vertical dashed lines represent the values of $x^{\pm}$ of \eqref{frequency-maximum-Kerr-ellipticity-angles-1} for each example.}
\label{plot-Kerr-ellipticity-angle}%
\end{figure}

The latter represents a frequency-dependent ellipticity Kerr angle (a consequence of non-null magnetic conductivity $\Sigma$), which is illustrated in Fig.~\ref{plot-Kerr-ellipticity-angle} in terms of the dimensionless parameter $x=\omega/\Sigma$. The maximum value of $\eta_{Kerr}=\pm \pi/4$ happens for specific values of frequency, given by 
	\begin{align}
		x^{\pm} &= \frac{\mu_{2} \sqrt{\mu_{1} \epsilon_{1}}}{\pm \mu_{1} \epsilon_{2} \mp \mu_{2} \epsilon_{1}}. \label{frequency-maximum-Kerr-ellipticity-angles-1}
	\end{align}
	The plus (minus) sign of $\eta_{Kerr}$ indicates the handedness of the reflected wave, which will be left- (right-) handed elliptically polarized \cite{Zangwill}. The plus (minus) sign of $x^{\pm}$ corresponds to the cases where $\mu_{1}\epsilon_{2} - \mu_{2} \epsilon_{1}$ is positive (negative), which leads to $\eta_{Kerr} >0$ ($\eta_{Kerr}<0$). Furthermore, the possible maximum values for Kerr ellipticity are $\pm \pi/4$ regardless of the values of permeabilities ($\mu_{1}, \mu_{2}$) and permittivities ($\epsilon_{1}, \epsilon_{2}$). In the special case of $\mu_{1}\epsilon_{2} - \mu_{2} \epsilon_{1} =0$, one finds $\eta_{Kerr} \rightarrow 0$, meaning that the reflected wave will be linearly polarized.

In the very low-frequency regime, the Kerr ellipticity angle (\ref{reflection-with-magnetic-conductivity-37-0}) behaves linearly with the frequency,
\begin{align}
\left. \eta_{Kerr} \right|_{low} \simeq \frac{\mu_{1} \epsilon_{2} -\mu_{2} \epsilon_{1} }{\mu_{2} \sqrt{\mu_{1}\epsilon_{1}}} \frac{\omega}{\Sigma} , \label{Kerr-ellipticiy-angle-low-frequency-1}
\end{align}
while for very high frequencies it goes as 
\begin{align}
\left. \eta_{Kerr} \right|_{high} \simeq \frac{ \mu_{2} \sqrt{\mu_{1}\epsilon_{1}}}{ \mu_{1} \epsilon_{2}-\mu_{2}\epsilon_{1}} \frac{\Sigma}{\omega} . \label{Kerr-ellipticiy-angle-low-frequency-1}
\end{align}
We point out that the general frequency behavior of the Kerr ellipticity (\ref{reflection-with-magnetic-conductivity-37-0}), depicted in Fig.~\ref{plot-Kerr-ellipticity-angle}, is analogous to the frequency-dependent Kerr effect observed in Weyl semimetals \cite{Sonowal}, where the optical activity is caused by the axion term. The evaluation of such an optical quantity in exotic materials endowed with magnetic conductivity may work as a tool to measure the magnitude of $\Sigma$.

\section{\label{Final-Remarks-section}Final Remarks}

In this work, we have analyzed the reflection at the interface between an ordinary dielectric and a dielectric endowed with magnetic conductivity, starting from the Fresnel coefficients for $s$ and $p$ polarizations, presented in Eqs. (\ref{reflection-with-magnetic-conductivity-6}) and (\ref{reflection-with-magnetic-conductivity-7}), respectively.

For both polarizations, two frequency-dependent reflection coefficients were obtained, $R^{\pm}$, in considering the refractive indices $n_{+}$ and $n_{-}$. The behavior of the reflection coefficients was examined in terms of the incidence angle and frequency. The dispersive character is a novelty that brings several interesting new reflection properties.

For the polarization-s, the two reflection coefficients are $R^{\pm}_{\perp}$.  As for $R^{+}_{\perp}$, its magnitude decreases monotonically with the frequency, as depicted in Fig. \ref{plot-R-perpendicular-mais-angulos-fixos} and Fig.~\ref{plot-R-perpendicular-mais-angulos-fixos-with-USUAL-cases}. Its behavior with the incidence angle $\theta_{1}$ is shown in Fig.~\ref{plot-R-perpendicular-mais-frequencia-fixa}, for a few values of $\Sigma$. All the coefficients $R^{+}_{\perp}$ increase continuously with the angle, while there occurs an overall magnitude enhancement with the value of $\Sigma$.

Concerning the reflection coefficient $R^{-}_{\perp}$, its behavior in terms of frequency is shown in Figs.~\ref{plot-R-perpendicular-menos-angulos-fixos} and  \ref{plot-R-perpendicular-menos-angulos-fixos-with-USUAL-cases}, being  substantially distinct from the one of $R^{+}_{\perp}$. For non-null $\Sigma$ values, the coefficient $R^{-}_{\perp}$ becomes equal to 1 in the low-frequency interval $0< \omega \leq \omega_{i}$, with $\omega_{i}$ given by \eqref{reflection-with-magnetic-conductivity-14}, representing a scenario of total reflection induced by the dependence on the frequency of $n_{2-}$. Indeed, for $0< \omega \leq \omega_{i}$, there occurs $n_{2-}<n_{1}$, and the Snell's law explains the total reflection. An additional interesting property is the fact of $R^{-}_{\perp}=0$ for a specific value of frequency, $\omega_{0}$, given by \eqref{reflection-with-magnetic-conductivity-16}, as shown in Figs.~\ref{plot-R-perpendicular-menos-angulos-fixos} and \ref{plot-R-perpendicular-menos-angulos-fixos-with-USUAL-cases}, representing a partial Breswter angle for this polarizarion. The behavior of $R^{-}_{\perp}$ in terms of the incidence angle $\theta_{1}$ was also examined in Fig.~\ref{plot-R-perpendicular-menos-frequencia-fixa}, where regions of total reflection, $R^{-}_{\perp}=1$, were reported for $\Sigma /\omega >1$, being defined for $\bar{\theta}_{i} < \theta_{1} < \pi/2$, with  $\bar{\theta}_{i}$ given by \eqref{reflection-with-magnetic-conductivity-20}. This total reflection region is also a consequence of the dispersive behavior of $R^{-}_{\perp}$, which does not occur in a conventional dielectric (see the dashed black line in Fig.~\ref{plot-R-perpendicular-menos-frequencia-fixa}).

For incident $p$-polarization, the reflection coefficients $R_{\parallel}^{\pm}$ are given in \eqref{reflection-with-magnetic-conductivity-9}. Figures \ref{plot-R-paralelo-mais-angulos-fixos} and  \ref{plot-R-paralelo-menos-angulos-fixos-modelo-2} depict the reflectance $R_{\parallel}^{+}$ and $R_{\parallel}^{-}$, respectively,  in terms of the frequency, exhibiting behavior similar to the ones of s-polarization (in terms of the frequency). The behavior of $R_{\parallel}^{\pm}$ with the incidence angle is illustrated in Figs.~\ref{plot-R-paralelo-mais-frequencia-fixa} and \ref{plot-R-paralelo-menos-frequencia-fixa}. The coefficient $R_{\parallel}^{+}$, shown in Fig. \ref{plot-R-paralelo-mais-frequencia-fixa}, initially decreases with $\theta_{1}$, reaching a null value at the Brewster angle, and so augmenting its magnitude to the maximum value ($R_{\parallel}^{+}=1$). It differs from the $R^{+}_{\perp}$ behavior of Fig. \ref{plot-R-perpendicular-mais-frequencia-fixa}), which rises up monotonically. In the case of $R_{\parallel}^{-}$, its behavior with $\theta_{1}$, given in Fig. \ref{plot-R-paralelo-menos-frequencia-fixa}, is nearly close to the profile of $R^{-}_{\perp}$ of Fig. \ref{plot-R-perpendicular-menos-frequencia-fixa}, both exhibiting a window of total reflection, $\bar{\theta}_{i} <\theta_{1} < \pi/2$, with $\bar{\theta}_{i}$ given by \eqref{reflection-with-magnetic-conductivity-20}.

The total reflection phenomenon can occur at the interface of Fig.~\ref{figura-figura-campos-incidente-refletido-transmitido} depending on the relations among the electromagnetic quantities describing the system. Each propagating mode can undergo total reflection according to the conditions presented in Tab.~\ref{tab:conditions-for-total-reflection}, revealing that non-null magnetic conductivity increases the possibilities for this phenomenon to occur.

As for critical angles for null reflection, the general conditions for $R^{+, -}_{\perp, \parallel}=0$ are presented in Tabs.~\ref{tab:conditions-for-critical-angles-R-perpendicular-1}, \ref{tab:conditions-for-critical-angles-R-paralelo-mais}, \ref{tab:conditions-for-critical-angles-R-paralelo-menos}, and \ref{tab:conditions-for-critical-angles-R-perpendicular-2}. For $s$-polarized incident wave, the same set of constitutive parameters does not yield simultaneously $R_{\perp}^{+} =0$ and $R_{\perp}^{-} = 0$. In this case, one has a specific critical angle at which $R_{\perp}^{+} =0$, $R_{\perp}^{-} \neq 0$ and another angle for $R_{\perp}^{-}=0$, $R_{\perp}^{+} \neq 0$. This is illustrated in Fig.~\ref{plot-critical-angles-R-perpendicular-exemplo-1}. It defines the partial Brewster angles $\theta_{s\pm}^{c}$ of \eqref{reflection-with-magnetic-conductivity-21}. On the other hand, for $p$-polarized incident wave, the critical angles for null reflection are given by \eqref{reflection-with-magnetic-conductivity-23}. Considering the particular conditions adopted in rows $c)$ of Tab. \ref{tab:conditions-for-critical-angles-R-paralelo-mais} and row $d)$ of Tab. \ref{tab:conditions-for-critical-angles-R-paralelo-menos}, one can find a total Brewster angle $\left. \theta_{p} \right|_{\bar{\omega}}$, with $\bar{\omega}$ given by \eqref{reflection-with-magnetic-conductivity-24}. Also, from Fig.~\ref{plot-critical-angles-R-paralelo-exemplo-1}, we observe a total Brewster angle of $\pi/4$.  Hence, a mixed incident wave with $s$ and $p$ polarization components becomes $s$-polarized upon incidence on the total Brewster angle $\left. \theta_{p} \right|_{\bar{\omega}}=\pi/4$.

We have also shown that the GH displacement for each propagating mode is shifted from the standard cases, as illustrated in Figs.~\ref{plot-Goos-Hanchen-s-polarization} and \ref{plot-Goos-Hanchen-p-polarization}. The magnetic conductivity allows angles $\bar{\theta}^{\parallel, \perp}_{-}$ for total reflection smaller than the usual scenario, yielding GH blue lines on the left of the usual GH black curves in Figs.~\ref{plot-Goos-Hanchen-p-polarization} and \ref{plot-Goos-Hanchen-s-polarization}. Furthermore, the conductivity $\Sigma$ allows angles $\bar{\theta}^{\parallel, \perp}_{+}$ for total reflection greater than the usual scenario, yielding the GH red lines on the right of the GH black lines in Figs.~\ref{plot-Goos-Hanchen-p-polarization} and \ref{plot-Goos-Hanchen-s-polarization}.

Another optical reflection signature of dielectric endowed with magnetic conductivity is the complex Kerr rotation, used to examine the reflected wave polarization upon normal incidence. We find a null Kerr rotation angle ($\theta_{Kerr}=0$), while the Kerr ellipticity angle $\eta_{Kerr}$ is given by \eqref{reflection-with-magnetic-conductivity-37-0}, whose frequency-dependent behavior of $\eta_{Kerr}$ is depicted in Fig.~\ref{plot-Kerr-ellipticity-angle}. An analogous dependence is also observed in Weyl semimetals, where the Kerr rotation is caused by axion terms. Thus, such reflection properties may provide a useful tool to optically probe the electromagnetic properties of chiral dielectric systems.

\appendix

\section{\label{Tables-conditions-for-null-reflection}General conditions for null reflection}

 \begin{table}[H]
 \centering
\caption{\small{General conditions for null $R_{\perp}^{+}$.}}
\begin{tabular}{p{.5cm} p{7.6cm}           } 
\hline \hline 
&\hspace{2.6cm} $R_{\perp}^{+}$  \\ [0.6ex]
\hline
\\
 a) &$\mu_{2} >\mu_{1}$, \, $\epsilon_{2} > \epsilon_{1}$,  $ \mu_{2} \epsilon_{1} > \mu_{1} \epsilon_{2}$,  $ \frac{\Sigma}{\omega} < \frac{    \left| \mu_{1}\epsilon_{2} - \mu_{2} \epsilon_{1} \right| }{      \mu_{2} \sqrt{ \mu_{1} \epsilon_{1}}} $                 \\ [1.3ex]
\colrule \\ 
b) & $\mu_{2} > \mu_{1}$, \, $\epsilon_{2} < \epsilon_{1}$, \, \newline $ \left( \frac{}{} \mu_{2} \epsilon_{2} <   \mu_{1} \epsilon_{1},     \frac{   \left|     \mu_{1}\epsilon_{1} - \mu_{2} \epsilon_{2} \right|  }{       \mu_{2} \sqrt{ \mu_{1} \epsilon_{1}}} < \frac{\Sigma}{\omega} < \frac{   \left| \mu_{1}\epsilon_{2} - \mu_{2} \epsilon_{1} \right|  }{    \mu_{2} \sqrt{ \mu_{1} \epsilon_{1}} } \right) $    or        $ \left( \mu_{2} \epsilon_{2} \geq \mu_{1} \epsilon_{1},   \frac{\Sigma}{\omega} <   \frac{  \left| \mu_{1} \epsilon_{2} - \mu_{2} \epsilon_{1} \right|  }{   \mu_{2} \sqrt{ \mu_{1} \epsilon_{1}}} \right)$                   \\ [1.3ex]
\colrule \\ 
 c)  &  $\mu_{2} > \mu_{1}$, \, $\epsilon_{2} =\epsilon_{1}$, \, $ \frac{\Sigma}{\omega} <   \frac{    \left| \mu_{1}\epsilon_{1} - \mu_{2} \epsilon_{2} \right|  }{  \mu_{2} \sqrt{ \mu_{1} \epsilon_{1}}} $         
\\ [1.3ex]
\colrule \\ 
d) &  $\mu_{2} < \mu_{1}$ , \, $  \epsilon_{2} < \epsilon_{1} $, \, $ \left(\frac{}{} \mu_{2} \epsilon_{1} \leq  \mu_{1} \epsilon_{2}  \right.$,      $ \left. \frac{\Sigma}{\omega} <  \frac{   \left|  \mu_{1}\epsilon_{1} - \mu_{2} \epsilon_{2}  \right| }{  \mu_{2} \sqrt{ \mu_{1} \epsilon_{1}}} \right) $   or    $ \left( \frac{}{}  \mu_{2}  \epsilon_{1} >  \mu_{1} \epsilon_{2} \right.$,        $  \left. \frac{      \left|  \mu_{1}\epsilon_{2} - \mu_{2} \epsilon_{1} \right|  }   {    \mu_{2}   \sqrt{  \mu_{1} \epsilon_{1}}}   <   \frac{\Sigma}{\omega}  <      \frac{     \left|  \mu_{1} \epsilon_{1} - \mu_{2} \epsilon_{2}  \right|   }{   \mu_{2} \sqrt{ \mu_{1} \epsilon_{1}  }}  \right) $     \\ [1.3ex]
\colrule \\ 
 e) &  $\mu_{2} < \mu_{1}$,  \,  $\epsilon_{2} = \epsilon_{1}$, \,   $ \frac{\Sigma}{\omega} <   \frac{   \left|  \mu_{1} \epsilon_{1} - \mu_{2} \epsilon_{2} \right|  }{   \mu_{2}  \sqrt{ \mu_{1} \epsilon_{1} }} $   \\ [1.3ex]
\colrule \\ 
 f)  & $\mu_{2} < \mu_{1}$, \, $\epsilon_{2} > \epsilon_{1}$, \, $ \mu_{2} \epsilon_{2} < \mu_{1} \epsilon_{1} $,         $\frac{\Sigma}{\omega} <   \frac{  \left|  \mu_{1}\epsilon_{1} - \mu_{2} \epsilon_{2}   \right| }{   \mu_{2}  \sqrt{ \mu_{1} \epsilon_{1}}} $       
\\ [1.3ex]
\hline \hline
\end{tabular}
\label{tab:conditions-for-critical-angles-R-perpendicular-1}
\end{table}

 \begin{table}[H]
 \centering
\caption{\small{General conditions for null $R_{\parallel}^{+}$.}}
\begin{tabular}{p{.5cm} p{7.6cm}           } 
\hline \hline 
&\hspace{2.6cm} $R_{\parallel}^{+}$  \\ [.6ex]
\hline
\\
 a) & $\mu_{2} >\mu_{1}$, \, $\epsilon_{2} > \epsilon_{1}$,  $ \left( \frac{}{} \mu_{2} \epsilon_{1} \leq  \mu_{1} \epsilon_{2}, \,  \frac{\Sigma}{\omega} >0  \right)  $          or     $ \left( \frac{}{} \mu_{2} \epsilon_{1} > \mu_{1} \epsilon_{2}, 
\frac{\Sigma}{\omega} >   \frac{  \left|  \mu_{2} \epsilon_{1}-\mu_{1} \epsilon_{2} \right| }{     \mu_{2} \sqrt{ \mu_{1} \epsilon_{1}}}  \right)$            \\ [1.3ex]
\colrule \\
b)  &       $\mu_{2} > \mu_{1}$,  \, $\epsilon_{2} < \epsilon_{1}$ , \, $\mu_{2}\epsilon_{2} <  \mu_{1} \epsilon_{1}  $, \newline \newline
$\frac{\Sigma}{\omega} <   \frac{   \left| \mu_{2}\epsilon_{2} - \mu_{1} \epsilon_{1} \right|  }{   \mu_{2} \sqrt{   \mu_{1} \epsilon_{1}}} $, \quad $\frac{\Sigma}{\omega} >  \frac{  \left| \mu_{2}\epsilon_{1}-\mu_{1} \epsilon_{2}\right| }{  \mu_{2} \sqrt{ \mu_{1} \epsilon_{1}}} $          \\ [1.3ex]
\colrule \\
 c)  &  $\mu_{2} > \mu_{1}$, \, $\epsilon_{2} = \epsilon_{1}$, \, $\frac{\Sigma}{\omega} >  \frac{  \left| \mu_{2}\epsilon_{2} - \mu_{1} \epsilon_{1} \right|  }{  \mu_{2} \sqrt{ \mu_{1} \epsilon_{1}}} $ 
\\ [1.3ex]
\colrule \\
d) &  $\mu_{2} < \mu_{1}$,  \, $\epsilon_{2} < \epsilon_{1}$, \, $\mu_{2} \epsilon_{1} \leq \mu_{1} \epsilon_{2}$, 
$\frac{\Sigma}{\omega} >  \frac{   \left| \mu_{2}\epsilon_{2} - \mu_{1}\epsilon_{1}\right|  }{  \mu_{2} \sqrt{ \mu_{1} \epsilon_{1}}} $    \\ [1.3ex]
\colrule \\ 
e) & $\mu_{2} < \mu_{1}$, \, $\epsilon_{2} > \epsilon_{1}$, \, $ \left( \mu_{2} \epsilon_{2} \geq \mu_{1}\epsilon_{1}, \,  \frac{\Sigma}{\omega} >0 \right)$    or   $ \left(\mu_{2} \epsilon_{2} < \mu_{1} \epsilon_{1}, 
 \frac{\Sigma}{\omega} >  \frac{  \left| \mu_{2}\epsilon_{2} - \mu_{1}\epsilon_{1}\right| }{  \mu_{2} \sqrt{ \mu_{1} \epsilon_{1}}}  \right)$   \\ [1.3ex]
\colrule \\ 
  f)  & $\mu_{2} < \mu_{1}$, \, $\epsilon_{2} < \epsilon_{1}$, \, $\mu_{2} \epsilon_{1} > \mu_{1} \epsilon_{2}$, \newline \newline
  $ \frac{\Sigma}{\omega} < \frac{  \left|  \mu_{2} \epsilon_{1} - \mu_{1} \epsilon_{2} \right|  }{  \mu_{2} \sqrt{ \mu_{1} \epsilon_{1}}}$ , \quad $\frac{\Sigma}{\omega} >  \frac{  \left| \mu_{2} \epsilon_{2} - \mu_{1}\epsilon_{1}\right| }{  \mu_{2} \sqrt{ \mu_{1} \epsilon_{1}}} $      \\ [1.3ex]
\colrule \\ 
g) & $\mu_{2} < \mu_{1}$ , \, $\epsilon_{2} =\epsilon_{1}$, \, $\frac{\Sigma}{\omega} >  \frac{ \left| \mu_{2} \epsilon_{2} - \mu_{1} \epsilon_{1} \right|  } {  \mu_{2} \sqrt{ \mu_{1} \epsilon_{1}}} $  \\ [1.3ex]
\colrule \\ 
h) & $\mu_{2} =\mu_{1}$ , \, $\left( \epsilon_{2} \geq \epsilon_{1}\right.$, $\left.\frac{\Sigma}{\omega} > 0 \right)$ or $ \left( \epsilon_{2} < \epsilon_{1}\right. $,
$ \frac{ \Sigma}{\omega} < \frac{  \left| \mu_{2} \epsilon_{2} - \mu_{1} \epsilon_{1} \right| }{ \mu_{2} \sqrt{ \mu_{1} \epsilon_{1}}}$,  \, $ \left. \frac{ \Sigma}{\omega} >  \frac{  \left| \mu_{2} \epsilon_{2} - \mu_{1} \epsilon_{1} \right| }{ \mu_{2} \sqrt{ \mu_{1} \epsilon_{1}}}  \right)$
\\ [1.3ex]
\hline \hline
\end{tabular}
\label{tab:conditions-for-critical-angles-R-paralelo-mais}
\end{table}

\begin{table}[h]
\centering
\caption{\small{General conditions for null $R_{\perp}^{-}$.}}
\begin{tabular}{p{.5cm} p{7.6cm}             } 
\hline \hline 
& \hspace{2.6cm} $ R_{\perp}^{-}$  \\ [0.6ex]
\hline
\\
a)  &       $\mu_{2} >\mu_{1}$, \, $\epsilon_{2} > \epsilon_{1}$, \,  $\left( \frac{}{} \mu_{2}  \epsilon_{1} \geq  \mu_{1} \epsilon_{2},  \frac{\Sigma}{\omega} <   \frac{  \left| \mu_{1} \epsilon_{1} - \mu_{2} \epsilon_{2} \right|  }{   \mu_{2} \sqrt{ \mu_{1} \epsilon_{1}}} \right) $      or  $ \left(  \frac{}{} \mu_{2}  \epsilon_{1} <  \mu_{1} \epsilon_{2}, \,   \frac{    \left| \mu_{1}\epsilon_{2} - \mu_{2} \epsilon_{1} \right|  }{      \mu_{2} \sqrt{\mu_{1} \epsilon_{1}}} < \frac{\Sigma}{\omega} <  \frac{   \left| \mu_{1}\epsilon_{1} - \mu_{2} \epsilon_{2} \right|  }{    \mu_{2} \sqrt{\mu_{1} \epsilon_{1}}} \right)$       \\ [1.3ex]
\colrule \\ 
b)                 &   $\mu_{2} >\mu_{1}$, \, $\epsilon_{2} < \epsilon_{1}$, \, $\mu_{2} \epsilon_{2} > \mu_{1} \epsilon_{1} $,   $\frac{\Sigma}{\omega} <  \frac{    \left| \mu_{1}\epsilon_{1}- \mu_{2} \epsilon_{2} \right| }{  \mu_{2}  \sqrt{\mu_{1} \epsilon_{1}}} $
\\ [1.3ex]
\colrule \\ 
 c)      &         $\mu_{2} > \mu_{1}$, \, $\epsilon_{2} =\epsilon_{1}$, \, $ \frac{\Sigma}{\omega} <   \frac{    \left| \mu_{1}\epsilon_{1} - \mu_{2} \epsilon_{2} \right|  }{  \mu_{2} \sqrt{ \mu_{1} \epsilon_{1}}} $
\\ [1.3ex]
\colrule \\ 
 d)        &       $\mu_{2} < \mu_{1}$, \, $\epsilon_{2} \leq \epsilon_{1}$  \, $\mu_{2} \epsilon_{1} < \mu_{1} \epsilon_{2}  $,   $\frac{\Sigma}{\omega} <  \frac{      \left| \mu_{1} \epsilon_{2} - \mu_{2} \epsilon_{1} \right| }{     \mu_{2} \sqrt{ \mu_{1} \epsilon_{1} }} $ 
\\ [1.3ex]
\colrule \\
 e)    &  $\mu_{2} < \mu_{1}$, \, $\epsilon_{2}  > \epsilon_{1}$, \, $ \left( \mu_{2} \epsilon_{2} \leq  \mu_{1} \epsilon_{1},      \frac{\Sigma}{\omega} <    \frac{   \left| \mu_{1} \epsilon_{2} - \mu_{2} \epsilon_{1}  \right| }{  \mu_{2} \sqrt{ \mu_{1} \epsilon_{1}}}  \right) $   or    $\left( \mu_{2} \epsilon_{2} >  \mu_{1} \epsilon_{1},     \frac{     \left| \mu_{1}\epsilon_{1} - \mu_{2} \epsilon_{2} \right|  }   {   \mu_{2}   \sqrt{ \mu_{1} \epsilon_{1}}}   <   \frac{\Sigma}{\omega}  <       \frac{     \left| \mu_{1} \epsilon_{2} - \mu_{2} \epsilon_{1}  \right|  }{     \mu_{2}   \sqrt{ \mu_{1} \epsilon_{1}  }} \right)  $  
\\ [1.3ex]
\colrule \\ 
 f)        &   $\mu_{2} < \mu_{1}$, \, $ \epsilon_{2} =\epsilon_{1}$, \,  $\frac{\Sigma}{\omega} <  \frac{    \left| \mu_{1} \epsilon_{2} - \mu_{2} \epsilon_{1} \right| }{   \mu_{2}  \sqrt{ \mu_{1} \epsilon_{1} }} $ \\[1.3ex]
\hline \hline
\end{tabular}
\label{tab:conditions-for-critical-angles-R-perpendicular-2}
\end{table}

\begin{table}[H]
\centering
\caption{\small{General conditions for null $R_{\parallel}^{-}$.}}
\begin{tabular}{p{.5cm} p{7.6cm}             } 
\hline \hline 
& \hspace{2.6cm} $ R_{\parallel}^{-}$  \\ [0.6ex]
\hline
\\	
a)  &       $\mu_{2} >\mu_{1}$, \, $\epsilon_{2}>\epsilon_{1}$, \, $\mu_{2} \epsilon_{1}< \mu_{1}\epsilon_{2}$, \newline
$\frac{\Sigma}{\omega} <  \frac{  \left| \mu_{2}\epsilon_{1}-\mu_{1} \epsilon_{2} \right| }{   \mu_{2} \sqrt{ \mu_{1}\epsilon_{1}}}$, \, $ \frac{\Sigma}{\omega} >   \frac{    \left| \mu_{2}\epsilon_{2} - \mu_{1}\epsilon_{1}\right|  }{ \mu_{2}  \sqrt{ \mu_{1}\epsilon_{1}}}$   \\ [1.3ex]
\colrule \\ 
b)            &    $\mu_{2} >\mu_{1}$, \, $\epsilon_{2} >\epsilon_{1}$, \, $\mu_{2} \epsilon_{1} \geq \mu_{1} \epsilon_{2}$, \, $\frac{\Sigma}{\omega} >  \frac{   \left| \mu_{2}\epsilon_{2} -\mu_{1}\epsilon_{1}\right| }{  \mu_{2} \sqrt{ \mu_{1}\epsilon_{1}}} $            \\ [1.3ex]
\colrule \\ 
c)                 &   $\mu_{2} >\mu_{1}$, \, $ \epsilon_{2} < \epsilon_{1} $, \, $ \left( \frac{}{} \mu_{2} \epsilon_{2} \leq\mu_{1} \epsilon_{1}, \, \frac{\Sigma}{\omega} > 0 \right)$  or  $ \left( \frac{}{} \mu_{2} \epsilon_{2} > \mu_{1} \epsilon_{1}, \,    \frac{\Sigma}{\omega} >    \frac{   \left| \mu_{1}\epsilon_{1} - \mu_{2} \epsilon_{2} \right| }{  \mu_{2} \sqrt{ \mu_{1} \epsilon_{1}}} \right)$
\\ [1.3ex]
\colrule \\
 d)        &       $\mu_{2} >  \mu_{1}$, \, $\epsilon_{2} = \epsilon_{1}$, \,    $\frac{\Sigma}{\omega} >   \frac{     \left| \mu_{2} \epsilon_{2} - \mu_{1} \epsilon_{1} \right| }{    \mu_{2} \sqrt{ \mu_{1} \epsilon_{1} }} $ 
\\ [1.3ex]
\colrule \\
 e)    &  $\mu_{2} < \mu_{1}$, \, $\epsilon_{2}  > \epsilon_{1}$, \, $ \left( \frac{}{} \mu_{2} \epsilon_{2} \leq  \mu_{1} \epsilon_{1},    \,  \frac{\Sigma}{\omega} >    \frac{  \left| \mu_{2} \epsilon_{1} - \mu_{1} \epsilon_{2}  \right| }{  \mu_{2}  \sqrt{\mu_{1} \epsilon_{1}}} \right)  $   or $ \left( \frac{}{} \mu_{2} \epsilon_{2} > \mu_{1} \epsilon_{1}, \newline \newline
  \frac{\Sigma}{\omega} <   \frac{   \left| \mu_{2} \epsilon_{2} - \mu_{1} \epsilon_{1} \right| }{   \mu_{2} \sqrt{ \mu_{1} \epsilon_{1} }}, \,  \frac{\Sigma}{\omega} > \frac{  \left| \mu_{2}\epsilon_{1} - \mu_{1} \epsilon_{2} \right|  }{\mu_{2} \sqrt{ \mu_{1} \epsilon_{1}}} \right)$
\\ [1.3ex]
\colrule \\ 
  f)     &    $\mu_{2} < \mu_{1}$, \, $ \epsilon_{2} =\epsilon_{1}$, \,   $   \frac{\Sigma}{\omega}  >  \frac{    \left| \mu_{2} \epsilon_{2} - \mu_{1} \epsilon_{1}  \right|  }{    \mu_{2}  \sqrt{ \mu_{1} \epsilon_{1}  }}  $  
  \\ [1.3ex]
\colrule \\
g) & $\mu_{2} < \mu_{1}$, \, $\epsilon_{2} <\epsilon_{1}$, $ \left( \frac{}{} \mu_{2} \epsilon_{1} < \mu_{1} \epsilon_{2}, 
\frac{\Sigma}{\omega} >  \frac{ \left| \mu_{2} \epsilon_{1} - \mu_{1} \epsilon_{2} \right|  }{  \mu_{2} \sqrt{ \mu_{1} \epsilon_{1}}} \right) $  or $ \left(\mu_{2} \epsilon_{1} \geq \mu_{1} \epsilon_{2}, \, \frac{\Sigma}{\omega} >0 \right)$ 
  \\ [1.3ex]
\colrule \\
h) & $\mu_{2} = \mu_{1}$, \,   $ \left( \frac{}{} \epsilon_{2} > \epsilon_{1}, \,
\frac{\Sigma}{\omega} < \frac{  \left| \mu_{2}\epsilon_{2} - \mu_{1} \epsilon_{1}\right| }{  \mu_{2} \sqrt{ \mu_{1} \epsilon_{1}}}, \,      \frac{\Sigma}{\omega} >  \frac{  \left| \mu_{2}\epsilon_{2} - \mu_{1} \epsilon_{1}\right| }{ \mu_{2} \sqrt{ \mu_{1} \epsilon_{1}}} \right)$    or   $ \left( \frac{}{} \epsilon_{2} \leq \epsilon_{1}, \, \frac{\Sigma}{\omega} > 0 \right) $ \\ [1.3ex]
\hline \hline
\end{tabular}
\label{tab:conditions-for-critical-angles-R-paralelo-menos}
\end{table}

\end{document}